\documentclass[usenatbib, onecolumn]{mnras}

\usepackage[utf8]{inputenc}
\usepackage{amsmath}
\usepackage{tikz}
\usepackage{lmodern}

\newcommand{\dee}{\mathrm{d}}

\newcommand{\grad}{\vec \nabla}

\newcommand{\tj}[6]{\begin{pmatrix} {#1} & {#2} & {#3}\\ {#4} & {#5} & {#6}\end{pmatrix}}

\newcounter{sarrow}

\usetikzlibrary{decorations.pathmorphing,shapes}

\newcommand{\dd}[1]{\delta_{\mathrm{D}}^{(3)}\left(#1\right)}

\newcommand{\vk}{\vec k}
\newcommand{\vK}{\vec K}

\newcommand{\vq}{\vec q}
\newcommand{\vx}{\vec x}

\renewcommand{\vec}[1]{\boldsymbol{#1}}

\definecolor{myblue}{rgb}{0.1, 0.4, 0.5}
\definecolor{color1}{rgb}{0.5, 0.5, 0.4}
\definecolor{lightpurple}{HTML}{9999FF}
\definecolor{lightblue}{HTML}{66CCFF}
\definecolor{sweetblue}{HTML}{33CCFF}
\definecolor{waterblue}{HTML}{0099CC}
\definecolor{steelblue}{HTML}{6699CC}
\definecolor{kingsblue}{HTML}{1CBCAC}
\definecolor{forest}{HTML}{006175}
\definecolor{oldgreen}{HTML}{008E3E}
\definecolor{saddlebrown}{HTML}{BC4C1C}
\definecolor{ube}{HTML}{301728}

\newcommand{\lb}{\left(}
\newcommand{\rb}{\right)}

\newcommand{\vqa}{\mathbf{q}_1}
\newcommand{\vqb}{\mathbf{q}_2}
\newcommand{\vqc}{\mathbf{q}_3}
\newcommand{\vqd}{\mathbf{q}_4}

\newcommand{\vQ}{\mathbf{Q}}

\newcommand{\qa}{q_1}
\newcommand{\qb}{q_2}
\newcommand{\qc}{q_3}
\newcommand{\qd}{q_4}

\newcommand{\hqa}{\hat{q}_1}

\newcommand{\hqc}{\hat{q}_3}

\newcommand{\hQ}{\hat{Q}}

\newcommand{\ka}{k_1}
\newcommand{\kb}{k_2}
\newcommand{\kc}{k_3}
\newcommand{\kd}{k_4}
\newcommand{\Li}{L_i}
\newcommand{\La}{L_1}
\newcommand{\Lb}{L_2}
\newcommand{\Lc}{L_3}
\newcommand{\Mi}{M_i}
\newcommand{\Ma}{M_1}
\newcommand{\Mb}{M_2}
\newcommand{\Mc}{M_3}
\newcommand{\DLM}{D_{\La\Lb\Lc}^{\Ma\Mb\Mc}}
\newcommand{\GLM}{G_{\La\Lb\Lc}^{\Ma\Mb\Mc}}
\newcommand{\HL}{H_{\La\Lb}}
\newcommand{\Ya}[1]{Y_{\La\Ma}(#1)}
\newcommand{\Yb}[1]{Y_{\Lb\Mb}(#1)}
\newcommand{\Yc}[1]{Y_{\Lc\Mc}(#1)}

\newcommand{\tji}{\tj{1}{1}{\Li}{0}{0}{0}}

\newcommand{\tjabc}{\tj{\La}{\Lb}{\Lc}{\Ma}{\Mb}{\Mc}}
\newcommand{\tjabco}{\tj{\La}{\Lb}{\Lc}{0}{0}{0}}

\newcommand{\jo}[1]{j_{0}(#1)}
\newcommand{\ja}[1]{j_{\La}(#1)}
\newcommand{\jb}[1]{j_{\Lb}(#1)}

\newcommand{\nj}[9]{\begin{pmatrix} #1 & #2 & #3 \\ #4 & #5 & #6 \\ #7 & #8 & #9 \end{pmatrix}}
\newcommand{\njabc}{\nj{\La}{\Lb}{\Lc}{1}{1}{1}{1}{1}{1}}

\begin{document}

\title[POP Spectra]{Parity-Odd Power Spectra: Concise Statistics for Cosmological Parity Violation}
\author[Jamieson et al.]{
    Drew Jamieson$^{1}$\thanks{E-mail: jamieson@mpa-garching.mpg.de (DJ)},
    Angelo Caravano$^{2}$\thanks{E-mail: caravano@iap.fr (AC)},
    Jiamin Hou$^{3,4}$\thanks{E-mail: jiamin.hou@ufl.edu (JH)},
    Zachary Slepian$^{3}$\thanks{E-mail: zslepian@ufl.edu (ZS)},
    \& Eiichiro Komatsu$^{1,4,5}$
    \\
    $^{1}$Max-Planck-Institut für Astrophysik, Karl-Schwarzschild-Straße 1, 85748 Garching, Germany\\
    $^{2}$Institut d'Astrophysique de Paris, UMR 7095 du CNRS et de Sorbonne Universit\'e, 98 bis Bd Arago, 75014 Paris, France\\
    $^{3}$Department of Astronomy, University of Florida, Gainesville, FL 32611, USA\\
    $^{4}$Max-Planck-Institut f\"ur extraterrestrische Physik, Gießenbachstraße 1, 85748 Garching, Germany\\
    $^{5}$Ludwig-Maximilians-Universit\"{a}t M\"{u}nchen, Schellingstr. 4, 80799 M\"{u}nchen, Germany\\
    $^{6}$Kavli Institute for the Physics and Mathematics of the Universe (Kavli IPMU, WPI), University of Tokyo, Chiba 277-8582, Japan
}

\maketitle
\label{firstpage}

\begin{abstract}
    We introduce the Parity-Odd Power (POP) spectra, a novel set of observables for probing parity violation in cosmological $N$-point statistics. POP spectra are derived from composite fields obtained by applying nonlinear transformations, involving also gradients, curls, and filtering functions, to a scalar field. This compresses the parity-odd trispectrum into a power spectrum. These new statistics offer several advantages: they are computationally fast to construct, estimating their covariance is less demanding compared to estimating that of the full parity-odd trispectrum, and they are simple to model theoretically. We measure the POP spectra on simulations of a scalar field with a specific parity-odd trispectrum shape. We compare these measurements to semi-analytic theoretical calculations and find agreement. We also explore extensions and generalizations of these parity-odd observables.
\end{abstract}


\section{Introduction}
\label{sec:introduction}

Modern physics has advanced considerably by proposing theoretical symmetries and testing whether or not those symmetries are respected or violated in nature. A crucial breakthrough for the Standard Model of particle physics occurred with the surprising discovery of weak nuclear parity violation in 1957 \citep{PhysRev.104.254, PhysRev.105.1413, PhysRev.105.1415}. Parity is the negation of spatial coordinates: $\vx\rightarrow-\vx$, and its violation implies that the laws of physics distinguish between right-handed and left-handed chiralities. While parity is known to be maximally violated on small scales by weak nuclear interactions, the status of parity on cosmological scales remains an open question.

Recent investigations using the method proposed in \citet{Cahn:2021ltpa} and \citealt{Cahn:2020axu} have uncovered intriguing evidence of cosmological parity violation in the large-scale structure (LSS) of the Universe. Notably, analyses of the BOSS galaxy four-point correlation function detected parity violation with significance of up to 7$\sigma$~\citep{Hou:2022wfj, Philcox:2022hkh}. Confirmation of parity violation in the galaxy clustering would have profound implications. If the signal is primordial in origin, it is a form of primordial non-Gaussianity and would greatly inform our early-Universe inflationary models \citep{Bartolo:2004if}. Current cosmic microwave background (CMB) observations are consistent with parity conservation \citep{Philcox:2023ffy,Philcox:2023xxk}. However, since the CMB is measured on a 2D surface and is sensitive to different scales than the LSS, it is still unclear what this implies for the BOSS signal. The central challenge in the current BOSS analysis lies in robustly estimating the covariance \citep{Cahn:2021ltpa}, which can be influenced by both instrumental systematics and observational effects \citep{Hou:2022wfj} as well as the type of mocks used to calibrate or compute the covariance \citep{hou_covar, Philcox:2024mmz}. 

Next-generation 3D spectroscopic surveys such as DESI \citep{DESI:2016fyo,DESI:2024mwx}, Euclid \citep{Amendola:2016saw}, Subaru PFS \citep{PFSTeam:2012fqu}, Roman \citep{roman_wang}, and SPHEREx \citep{SPHEREx:2014bgr}, as well as photometric efforts such as LSST \citep{LSST:2008ijt}, will provide significantly more data than is currently available. To constrain large-scale parity violation robustly with this wealth of data, we need analysis tools that facilitate accurate covariance estimation and characterization of observational systematics.

For scalar fields like the primordial curvature perturbation or the matter density contrast, $N$-point statistics are sensitive to parity at orders four and above \citep{shiraishi:2016mok}. Measurements of these correlators have an enormous number of degrees of freedom (\textit{e.g.} the BOSS analysis of \citealt{Hou:2022wfj} had 18,000). Covariance estimation for such large data vectors requires a large number of simulated mock datasets, often combined with approximate analytic covariance templates (\textit{e.g.} \citealt{Hou:2021ncj}, also \citealt{SE_3pt}, \citealt{Xu_2012}). Even with statistically accurate covariance estimation, systematics in observational data can bias both the detected signal and its significance.

A complementary approach to high-order $N$-point statistics is lower-order statistics, such as power spectra, that are computed on composite fields. These composite fields are created by applying nonlinear transformations to the original scalar field. Composite-field statistics average or compress the high-dimensional correlators down to low-dimensional correlators. As a result, the covariances of these composite-field power spectra are less demanding to estimate, due to having fewer degrees of freedom, while still carrying higher-order information. The trade-off involved in this approach is the loss of information through compression. Examples of weighted, compressed analyses for bispectra and parity-even trispectra have been done before and are known as \emph{skew-spectra} and \emph{kurt-spectra}~\citep{Schmittfull:2014tca,Schmittfull:2020hoi,MoradinezhadDizgah:2019xun,Hou:2022rcd,Munshi:2021uwn}, as well as works on the unweighted integrated three-point correlation function \citep{se_boss_integ_3}, trispectrum \citep{gualdi_sims, gualdi_22}, and four-point function \citep{sabiu}; a Fourier-transform-based algorithm for the full four-point function is \cite{sunseri}. An estimator of the CMB lensing power spectrum can also be constructed in this manner \citep{Hu:2001fa}. 

In the current work, we introduce new methods for detecting cosmological parity violation by constructing power-spectrum-like statistics sensitive to parity. We call our new observables \emph{parity-odd power spectra}, or \emph{POP spectra}. The POP spectra are compressions of the six-dimensional four-point statistics down to one-dimensional power spectra. They are computationally efficient to construct and their lower dimensionality alleviates the burden of full four-point covariance estimation. 

Importantly, POP spectra are sensitive to soft limits of parity-violating trispectrum shapes, which can help to place strong constraints on primordial parity violation. Moreover, being formulated in Fourier space, they are easy to interpret and can facilitate theoretical understanding of the signal, which is crucial to distinguish it from potential observational systematics. It is also possible to construct real-space and spherical-harmonic-space versions of these statistics. As we will demonstrate below, constructing parity-odd statistics is nontrivial, requiring more sophisticated techniques than those required for parity-even compressed estimators.

This work is structured as follows. In \S\ref{sec:ppvt}, we discuss the primordial parity-odd trispectrum. In \S\ref{sec:pvps}, we present two POP spectra constructions based on composite vector and scalar fields. In \S\ref{sec:valid}, we validate the estimator on simulated data, including comparisons with semi-analytical calculations. In \S\ref{sec:extension}, we explore extensions and generalizations of our parity-odd observables. Finally, we conclude and discuss prospects for measuring these parity-odd statistics on observational data in \S\ref{sec:conclude}.

\section{Parity-Violating Trispectrum}
\label{sec:ppvt}

\begin{figure}
    \centering
    \includegraphics[width = 0.4\linewidth]{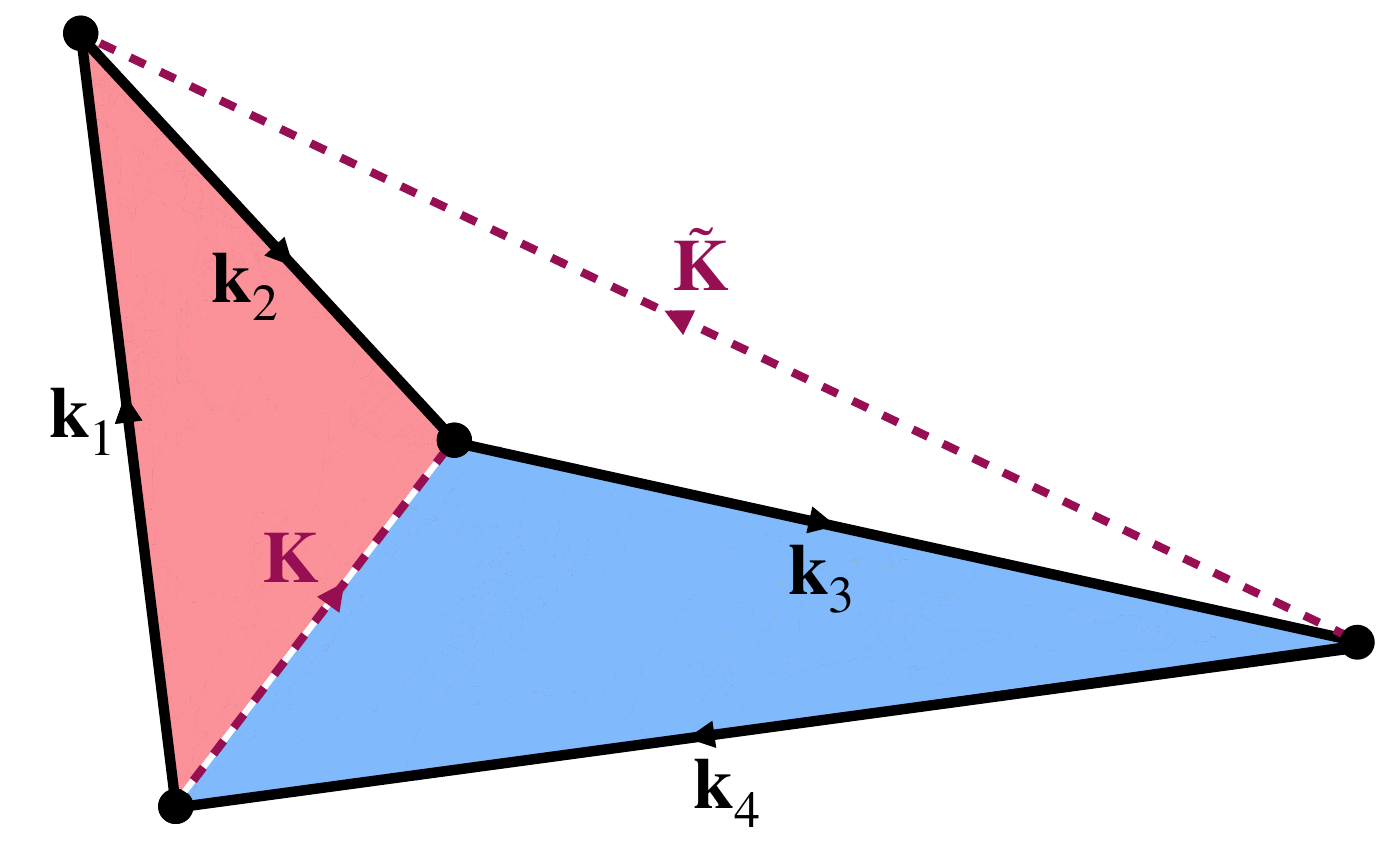}
    \caption{Here we display a Fourier-space tetrahedron representing the trispectrum configuration $T(\vk_1,\vk_2,\vk_3)$. The diagonals (in dashed magenta) are defined as $\vK=\vk_1+\vk_2$ and $\tilde \vK = \vk_1+\vk_4$. The tetrahedron is formed by joining two triangles, $\{\vk_1, \vk_2, -\vK \}$ in red and $\{\vk_3, \vk_4, \vK\}$ in blue, along their shared edge.}
    \label{fig:tet}
\end{figure}

    Under statistical homogeneity and isotropy, the four-point correlation function is the lowest-order parity-sensitive statistic for a single scalar field. In Fourier space, this corresponds to the trispectrum, $T(\vk_1, \vk_2, \vk_3)$, defined through the four-point correlator,
\begin{align} \label{eq:tspec}    
    \big\langle \Phi(\vk_1) \Phi(\vk_2) \Phi(\vk_3) \Phi(\vk_4) \big\rangle
    \equiv
    (2\pi)^3 \dd{\vk_1 + \vk_2 + \vk_3 + \vk_4} \, T(\vk_1, \vk_2, \vk_3) \, .
\end{align}
The Dirac delta function imposes translational invariance by requiring that the sum of the four wave vectors vanishes. Thus, they form a closed loop in Fourier space, and define a tetrahedron, as displayed in Fig.~\ref{fig:tet}. More precisely, the four wave vectors define an equivalence class of tetrahedra, consisting of the 24 tetrahedra produced by permuting them. These are not all geometrically distinct, since a cyclic permutation of wave vectors within a given tetrahedron yields a congruent tetrahedron. The three independent wave vectors, which we take to be $\vk_1$, $\vk_2$, and $\vk_3$, uniquely determine these tetrahedra. Statistical isotropy, however, reduces the dimensionality of the trispectrum from nine to six continuous degrees of freedom. We will choose the wave vector magnitudes $k_1$, $k_2$, $k_3$, and $k_4$ as four of these continuous parameters, and refer to them as \emph{sides} of the tetrahedron. Then we take $K = |\vk_1 + \vk_2| = |\vk_3 + \vk_4|$ and $\tilde{K} = |\vk_1 + \vk_4| = |\vk_2 + \vk_3|$ as the other two parameters and refer to these as \emph{diagonals} to distinguish them from the wave vector magnitudes.

Among the set of 24 tetrahedra, there is an additional discrete degree of freedom that splits them into two groups of 12.
These are distinguished by the sign of the vector triple product $\vk_1 \cdot (\vk_2 \times \vk_3)$. This can be thought of as the \emph{handedness} or helicity of the tetrahedron. The triple product is positive for a right-handed configuration and negative for a left-handed configuration. Under a parity transform, the sign of the triple product changes, so a parity transformation interchanges the two helicities.

We can isolate the trispectrum for only right-handed configurations, and obtain the right-handed trispectrum $T_{\rm R}(\vk_1, \vk_2, \vk_3)$, and similarly for the left-handed trispectrum $T_{\rm L}(\vk_1, \vk_2, \vk_3)$. Let $\pi_{123}\equiv{\rm sgn}\big(\vk_1 \cdot (\vk_2 \times \vk_3)\big)$, then
\begin{align}
    T(\vk_1, \vk_2, \vk_3) = \frac{1+\pi_{123}}{2} T_{\rm R}(\vk_1, \vk_2, \vk_3) + \frac{1-\pi_{123}}{2} T_{\rm L}(\vk_1, \vk_2, \vk_3) \, .
\end{align}
The right- and left-handed parts of the trispectrum are not necessarily equal. If they are different, this indicates that the trispectrum has a parity-odd component. For the modes of a real field, a parity transformation is complex conjugation,
\begin{align}
    \Phi(-\vk) = \Phi(\vk)^* \, ,
\end{align}
so the parity-even component of the trispectrum is its real part and is proportional to the sum of $T_{\rm R} + T_{\rm L}$. The parity-odd component is its imaginary part and is proportional to the difference $T_{\rm R} - T_{\rm L}$:
\begin{align}
    T(\vk_1, \vk_2, \vk_3) & = \frac{1}{2} \Big(T_{\rm R}(\vk_1, \vk_2, \vk_3) + T_{\rm L}(\vk_1, \vk_2, \vk_3) \Big) + \frac{\pi_{123}}{2} \Big(T_{\rm R}(\vk_1, \vk_2, \vk_3) - T_{\rm L}(\vk_1, \vk_2, \vk_3) \Big) \\
    & = \ T_+(\vk_1, \vk_2, \vk_3) + i T_-(\vk_1, \vk_2, \vk_3) \, .
\end{align}
The relationship between the right-/left-trispectra and the parity-even/parity-odd trispectra are,
\begin{align}
    T_{\rm R/L} = T_+ \pm i \pi_{123} T_- \, , \\
    T_{\pm} = \frac{1}{2}(-i\pi_{123})^{(1\mp1)/2} \big(T_{\rm R} \pm T_{\rm L} \big) \, .
\end{align}

Due to isotropy, the only vectors we can use to form a parity-odd structure are $\vk_1$, $\vk_2$, and  $\vk_3$, so $T_-(\vk_1, \vk_2, \vk_3)$ must be proportional to the triple product $\vk_1 \cdot (\vk_2 \times \vk_3)$. We parameterize the shape of the imaginary trispectrum as \citep{Coulton:2023oug}
\begin{align} \label{eq:tauminu}
    T_-(\vk_1, \vk_2, \vk_3) = \vk_1 \cdot (\vk_2 \times \vk_3)\, \tau_-(k_1, k_2, k_3, k_4, K, \tilde{K}) \, .
\end{align}
The left-hand side of Eq.~\eqref{eq:tspec} is totally symmetric under the interchange of any of the wave vectors, and the triple product in Eq.~\eqref{eq:tauminu} is totally antisymmetric. These symmetries require that $\tau_-(k_1, k_2, k_3, k_4, K, \tilde{K})$ is totally antisymmetric under the interchange of any two of the four wave vectors. The trispectrum cannot depend on one diagonal and not the other since $K\rightarrow\tilde{K}$ under the interchange of $\vk_2$ and $\vk_4$. Also, under the interchange of $\vk_2$ and $\vk_3$, $K$ becomes $|\vk_1 + \vk_3|$, which is not independent, since
\begin{align}
    |\vk_1 + \vk_3|^2 = -K^2 - \tilde{K}^2 + k_1^2 + k_2^2 + k_3^2 + k_4^2 \, .
\end{align}

This parameterization is not unique. We could choose any six geometrically independent quantities of the tetrahedra to parameterize the continuous degrees of freedom. For example, we could reparameterize by replacing $K$ with the angle between $\vk_1$ and $\vk_2$. Similarly, we can choose either parity or helicity to parameterize the discrete degree of freedom.

\section{Parity-Odd Power Spectra}
\label{sec:pvps}

CMB and LSS observables are consistent with cosmic structure originating from a single primordial scalar potential, $\Phi(\vx)$. In this section, we construct parity-odd power-spectrum-like observables, which we call POP spectra. These are compressions of the six-dimensional parity-odd trispectrum down to a one-dimensional power spectrum of composite fields. First, we construct the \emph{vector POP spectrum}, defined below in Eq. \eqref{eqn:P_vpv}, by cross-correlating a vector field with a pseudovector field, both derived as composite fields quadratic in $\Phi$. Next, we construct the \emph{scalar POP spectrum}, defined below in Eq. \eqref{eqn:P_sps}, by cross-correlating a scalar field with a pseudoscalar field. The latter is formed as a composite field that is cubic in $\Phi$. 

\subsection{Vector POP Spectrum}
\label{subsec:vpv}

Our goal in this subsection is to construct composite vector and pseudovector fields out of a scalar field such that the cross-power spectrum between them is a compressed estimator of the parity-odd trispectrum. Since a vector is parity odd and a pseudovector is parity even, their cross-correlation will be a POP spectrum. Obtaining a vector field from a scalar is trivial: take its gradient, $\grad\Phi(\vx)$. Obtaining a pseudovector is non-trivial. One possibility is to convolve the field with a filtering or smoothing function $f_{a}(\vx)$:
\begin{align}
    \Phi_{a}(\vx) = & \int \dee^3 y\, f_{a}(\vx - \vec{y}) \Phi(\vec{y}) \\
     = & \int_{\vk} f_{a}(\vk) \Phi(\vk) e^{i \vk\cdot\vx} \, ,
\end{align}
where the integral notation is defined in Table~\ref{tab:ints}. For two distinct filtering functions, $f_a(\vx)$ and $f_{b}(\vx)$, the composite vector field $\vec{V}_{ab}(\vx)\equiv\Phi_{a}(\vx)\grad\Phi_{b}(\vx)$ is not the gradient of a scalar, so it also has a curl component. Therefore, we can carry out a Helmholtz decomposition,
\begin{align}
\label{eq:helmholtz}
    \vec{V}_{ab}(\vx) = \grad \phi_{ab}(\vx) - \grad \times \vec{A}_{ab}(\vx) \, .
\end{align}
Since $\vec{V}_{ab}$ and the spatial derivatives are parity odd, $\vec{A}_{ab}$ is a pseudovector. Notice that if the two fields, $\Phi_{a}$ and $\Phi_{b}$, were identical, $\vec{A}_{ab}$ would vanish. Although one of these can be the original, unfiltered field. We impose that its divergence vanishes because any possible divergence would not contribute to $\vec{V}_{ab}(\vx)$. Then we solve for its modes by taking the curl of $\vec{V}_{ab}(\vx)$ and solving the resulting Poisson equation,
\begin{align} \label{eq:amodes}
    \vec{A}_{ab}(\vk) = -\frac{1}{k^2} i\vk \times \vec{V}_{ab}(\vk) \, .
\end{align}

In general, we are free to choose four distinct filter functions:  $f_a(\vx)$, $f_b(\vx)$, $f_c(\vx)$, and  $f_d(\vx)$. From the first two, we define the composite vector field $\vec{V}_{ab}(\vx)$. From the second two, we define the composite vector field $\vec{V}_{cd}(\vx)$ and then isolate the pseudovector field $\vec{A}_{cd}(\vx)$ from it. The dot product $\vec{V}_{ab}(\vx)\cdot\vec{A}_{cd}(\vx)$ is parity odd. As a result, we can define our first POP spectrum as 
\begin{align}
\label{eqn:P_vpv}
    \langle \vec{V}_{ab}(\vk) \cdot \vec{A}_{cd}(\vk') \rangle = (2\pi)^3 \dd{\vk+\vk'} P_{\mathrm{vector}}(k) \, .
\end{align}
 Explicitly, we can express the power spectrum as
\begin{align}
    \label{eq:vpv}
    P_{\mathrm{vector}}(k) = \frac{1}{\mathcal{N}_{\mathrm{vector}}(k)}
    \int\displaylimits_{\vq_1,\vq_2,\vq_3,\vq_4} \hspace{-8pt} \Big[
    (2\pi)^6 \dd{\vq_1 + \vq_2 - \vk} \, \dd{\vq_3 + \vq_4 + \vk} \, 
    f_{a}(\vq_1) f_{b}(\vq_2) f_{c}(\vq_3) f_{d}(\vq_4)  \nonumber \\ 
    \times \, \left[\vk \cdot (\vq_1 \times \vq_3)\right]^2 \, \tau_{-}(q_1, q_2, q_3, q_4, k, |\vq_1 + \vq_4|) \Big] \, .
\end{align}
The wave vector integrals are defined in Table~\ref{tab:ints}. The normalization factor $\mathcal{N}_\mathrm{vector}(k)$ is somewhat arbitrary, but if chosen poorly, the estimator will depend strongly on the geometry and resolution of the region where we sample $\Phi(\vx)$. A sensible choice is
\begin{align} 
    \label{eq:vpvnorm}
    \mathcal{N}_{\mathrm{vector}}(k) = \displaystyle\int\displaylimits_{\vq_1,\vq_2,\vq_3,\vq_4} (2\pi)^6 \dd{\vq_1 + \vq_2 - \vk} \, \dd{\vq_3 + \vq_4 + \vk} \, f_{a}(\vq_1) f_{b}(\vq_2) f_{c}(\vq_3) f_{d}(\vq_4) \, \left[\vk \cdot (\vq_1 \times \vq_3)\right]^2  \, ,
\end{align}
so that Eq. \eqref{eq:vpv} is a weighted average over the parity-odd trispectrum shape $\tau_{-}(q_1, q_2, q_3, q_4, k, |\vq_1 + \vq_4|)$. 

If we choose isotropic filtering functions that depend only on the magnitudes of the $\vq_i$, the vector POP spectrum is a real-valued quantity that is non-vanishing only if the parity-odd trispectrum is non-vanishing. With isotropic filtering functions, the parity-even part of the trispectrum does not contribute to the imaginary part of the POP spectrum. To see this, we note that the parity-even trispectrum is invariant under $\vq_i\rightarrow-\vq_i$ while the triple product receives a minus sign. Therefore, the integrand is antisymmetric and the integral vanishes.\footnote{The sign of the $\vk$ in the Dirac delta functions does not matter, since $P_{\rm{vector}}(k)$ is a function of only the magnitude of $\vec{k}$.}

The vector-pseudovector construction compresses the trispectrum down to a power spectrum by correlating two fields of order $\Phi^2$. The modes of these fields represent triangles of wave vectors satisfying $\vk + \vq_1 + \vq_2 = 0$ and $\vk' + \vq_3 + \vq_4 = 0$. Computing the power spectrum enforces that the two triangles connect along a side of equal length, $k$, which imposed $\vk + \vk' = 0$. Thus, the two triangles form a tetrahedron (shown in Fig.~\ref{fig:tet}) as the trispectrum requires. The vector POP spectrum is a weighted average over all trispectrum configurations while holding fixed the diagonal side length $k$, which is the argument of this POP spectrum. For this reason, the vector POP carries crucial information about the soft limit where the diagonal approaches zero.

\begin{table}
    \centering
    \begin{tabular}{|c@{\hspace{2em}}|@{\hspace{2em}}l|}
        \hline
        \textbf{Notation} & \textbf{Definition} \\
        \hline
        $\displaystyle\int_{\vk}$ & $\displaystyle\int \frac{\dee^3 k}{(2\pi)^3}$ \\
        \hline
        $\displaystyle\int\displaylimits_{\vq_1,...,\vq_n}$ & $\displaystyle \int_{\vq_1}...\int_{\vq_n}$ \\
        \hline
        $\displaystyle\int_q$ & $\displaystyle\int_0^{\infty} \frac{q^2 \dee q}{2 \pi^2}$ \\
        \hline
        $\displaystyle\int\displaylimits_{q_1,...,q_n}$ & $\displaystyle\int_{q_1} ... \int_{q_n}$ \\
        \hline
        \hline
    \end{tabular}
    \caption{Here, we list our Fourier-space integral notation and conventions. The bold symbols indicate wave vectors and the integral is over all of Fourier space. The nonbold symbols indicate the wave vector magnitude.}
    \label{tab:ints}
\end{table}

If the trispectrum has no explicit dependence on the diagonals and the filtering functions depend only on the magnitudes of the wave vectors in Fourier space, we can compute the angular integrals in Eqs.~(\ref{eq:vpv}--\ref{eq:vpvnorm}) analytically. Then the expression for the vector POP spectrum simplifies (as detailed in Appendix \ref{app:bin}) to
\begin{align}
    \label{eq:vpvth}
    P_{\mathrm{vector}}(k) = \frac{\pi^4}{2\mathcal{N}_{\mathrm{vector}}(k)}
    \int\displaylimits_{q_1,q_2,q_3,q_4} \Big[ 
    \frac{q_1}{q_2} \, \Theta\big(\sin^2\!\theta_{12}\big) \sin^2\!\theta_{12} \,
    \frac{q_3}{q_4} \, \Theta\big(\sin^2\!\theta_{34}\big) \sin^2\!\theta_{34} \, 
    f_{a}(q_1) f_{b}(q_2) f_{c}(q_3) f_{d}(q_4) \nonumber \\
    \times \, \tau_{-}(q_1, q_2, q_3, q_4) \Big] \, ,
\end{align}
and
\begin{align}
    \label{eq:vecnorm}
    \mathcal{N}_{\mathrm{vector}}(k) = \frac{\pi^4}{2} 
    \int\displaylimits_{q_1,q_2,q_3,q_4} \,
    \frac{q_1}{q_2} \, \Theta\big(\sin^2\!\theta_{12}\big) \sin^2\!\theta_{12} \,
    \frac{q_3}{q_4} \, \Theta\big(\sin^2\!\theta_{34}\big) \sin^2\!\theta_{34} \, 
    f_{a}(q_1) f_{b}(q_2) f_{c}(q_3) f_{d}(q_4) \, .
\end{align}
The radial Fourier integrals are defined in Table~\ref{tab:ints}. Here, $\Theta(x)$ is the Heaviside function,
\begin{align}
    \Theta(x) = 
    \begin{cases}
        0, & \text{if } x < 0 \\
        \displaystyle \frac{1}{2}, & \text{if } x = 0 \\
        1, & \text{if } x > 0
    \end{cases} ,
\end{align}
and $\theta_{12}$ the angle between $\vq_1$ and $\vk$ in the first triangle, so
\begin{align}
    \label{eq:sinsq12}
    \sin^2\theta_{12} = 1 - \left(\frac{q_1^2 + k^2 - q_2^2}{2 q_1 k}\right)^2 \, ,
\end{align}
and similarly for $\theta_{34}$, with 1, 2 $\rightarrow$ 3, 4. Notice that if $q_i$, $q_j$ and $k$ fail to satisfy the triangle inequalities $|q_j - q_j| \leq k \leq q_i + q_j$, then the right-hand sides in Eqs.~\eqref{eq:sinsq12} would be negative and the corresponding sine would be purely imaginary. Thus, the Heaviside functions in Eqs.~(\ref{eq:vpvth}--\ref{eq:vecnorm}) enforce the triangle inequalities. 

As a final remark, note that the pseudovector defined in Eq. \eqref{eq:amodes} is not unique. For example, one could alternatively take the following curl,
\begin{align} \label{eq:pvecb}
    \vec{B}_{ab}(\vx) = \grad \Phi_a(\vx) \times \grad \Phi_b(\vx) \, .
\end{align}
Using this instead of $\mathbf{A}_{ab}$ simply multiplies the vector POP spectrum $P_{\rm vector}(k)$ by a factor of $-k^2$. Instead of applying separate filtering functions to each field to the cross-product does not vanish, we could define an antisymmetric convolution kernel. Using convolutions kernel is similar to the skew spectrum \citep{Schmittfull:2014tca}, and is the method used for clustering fossil estimators \citep{Jeong:2012df}. These are better suited to model-specific optimal estimators, whereas in this work we are taking a more model-independent approach.

\subsection{Scalar POP Spectrum}
\label{subsec:sps}

We can construct another POP spectrum by cross-correlating a scalar field with a pseudoscalar field. The construction of a pseudoscalar, similar to that of the pseudovector from the previous subsection, is nontrivial and requires convolving the scalar field with filtering functions. In this case, we first construct the pseudovector given in Eq. \eqref{eq:pvecb}. We have the freedom to further smooth this pseudovector by another filtering function, with modes
\begin{align}
    \vec{B}_{bcd}(\vQ) = -f_d(\vQ) \int\displaylimits_{\vq_3,\vq_4} (2\pi)^3\dd{\vq_3 + \vq_4 - \vQ} f_b(\vq_3) f_c(\vq_4) \, (\vq_3 \times \vq_4) \Phi(\vq_3) \Phi(\vq_4) \, .
\end{align}
Then, we construct the triple product field
\begin{align}
    \Psi_{abcd}(\vx) = \grad \Phi_a(\vx) \cdot \vec{B}_{bcd}(\vx) \, ,
\end{align}
which is a pseudoscalar. Its cross-power spectrum with $\Phi(\vk)$ is a POP spectrum,
\begin{align}\label{eqn:P_sps}
    \langle \Phi(\vk) \Psi_{abcd}(\vk') \rangle = (2\pi)^3\dd{\vk+\vk'} P_{\mathrm{scalar}}(k) \, .
\end{align}
The scalar POP spectrum is given by
\begin{align} 
    \label{eq:sps}
    P_{\mathrm{scalar}}(k) = \frac{1}{\mathcal{N}_{\mathrm{scalar}}(k)}
    \int\displaylimits_{\vq_2,\vq_3,\vq_4,\vec Q} \Big[
    (2\pi)^6 \dd{\vk + \vq_2 + \vec Q} \, \dd{\vq_3 + \vq_4 - \vec Q} \, 
    f_{a}(\vq_2) f_{b}(\vq_3) f_{c}(\vq_4) f_{d}(\vec Q) \nonumber \\ 
    \times \left[\vec Q \cdot (\vq_2 \times \vq_4)\right]^2 \, 
    \tau_{-}(k, q_2, q_3, q_4, Q, |\vq_2 + \vq_3|) \Big]  \, ,
\end{align}
with normalization
\begin{align}
    \mathcal{N}_{\mathrm{scalar}}(k) =
 \int\displaylimits_{\vq_2,\vq_3,\vq_3,\vec Q} &
 (2\pi)^6 \dd{\vk + \vq_2 + \vec Q} \, \dd{\vq_3 + \vq_4 - \vec Q} \, 
 f_{a}(\vq_2) f_{b}(\vq_3) f_{c}(\vq_4) f_{d}(\vec Q) \left[\vec Q \cdot (\vq_2 \times \vq_4)\right]^2  \, .
\end{align}
As with the vector-pseudovector construction, we choose the normalization so that the POP spectrum is a weighted average over the imaginary trispectrum. In this case, we have constructed the compressed four-point statistic by cross-correlating a pseudoscalar of order $\Phi^3$ with the original potential $\Phi$. The resulting power spectrum is an average over the trispectrum tetrahedra, fixing the magnitude of one of the wave vector sides. The average integrates over the diagonal $\vQ$ this time (corresponding to the $k$ in Eqs.~\ref{eqn:P_vpv}--\ref{eq:vecnorm}), and the scalar POP spectrum is a function of only the fixed wave vector magnitude $k$, a side length of the tetrahedron. This property makes the scalar POP particularly useful for probing the soft limit where one of the side lengths approaches zero.

The side length that we label $k$ here was labelled $q_1$ for the vector POP spectrum in Eqs.~(\ref{eq:vpv}--\ref{eq:vecnorm}), so in the following equations a subscript 1 will refer to this side length. Again, if the trispectrum does not depend on the diagonals and the filtering functions depend only on the magnitudes of the wave vectors, we can compute the angular integrals analytically, and the expression simplifies (as detailed in Appendix \ref{app:bin}) to
\begin{align}
    \label{eq:spsth}
 P_{\mathrm{scalar}}(k) = \frac{\pi^4}{2k\mathcal{N}_{\mathrm{scalar}}(k)}
 \int\displaylimits_{q_2,q_3,q_3,Q} \Big[
 q_2\, \Theta\big(\sin^2\!\theta_{21}\big) \sin^2\!\theta_{21}\,
 \frac{q_4}{q_3} \, \Theta\big(\sin^2\!\theta_{43}\big) \sin^2\!\theta_{43} \, 
 f_{a}(q_2) f_{b}(q_3) f_{c}(q_4) f_{d}(Q) \nonumber \\
 \times \, \tau_{-}(k, q_2, q_3, q_4) \Big] \, ,
\end{align}
and
\begin{align}
    \label{eq:scalarnorm}
    \mathcal{N}_{\mathrm{scalar}}(k) = \frac{\pi^4}{2 k} \int\displaylimits_{q_2,q_3,q_3,Q} \,
    q_2 \, \Theta\big(\sin^2\!\theta_{21}\big) \sin^2\!\theta_{21} \,
    \frac{q_4}{q_3} \, \Theta\big(\sin^2\!\theta_{43}\big) \sin^2\!\theta_{43} \, 
    f_{a}(q_2) f_{b}(q_3) f_{c}(q_4) f_{d}(Q) \, .
\end{align}
Here, we evaluate $\sin^2 \theta_{21}$ and $\sin^2\theta_{43}$ using
\begin{align}
    \label{eq:cos21}
    \sin^2 \theta_{21} = 1 - \left(\frac{q_2^2 + Q^2 - k^2}{2 q_2 Q}\right)^2 \, , \\
    \sin^2 \theta_{43} = 1 - \left(\frac{q_4^2 + Q^2 - q_3^2}{2 q_4 Q}\right)^2 \, .
\end{align}

\subsection{Constructing the Estimators}
\label{subsec:est}

To construct the vector POP spectrum from a scalar field $\Phi(\vx)$ on a discrete grid, we start by taking its Fourier transform, $\Phi(\vq)$. Then, we make four copies of these modes and rescale them by the filter functions $\{f_{a}(\vq), f_{b}(\vq), f_{c}(\vq), f_{d}(\vq)\}$, obtaining $\{\Phi_a(\vq), \Phi_b(\vq), \Phi_c(\vq)$, $\Phi_d(\vq)\}$. Multiplying the modes of $\Phi_b(\vq)$ and $\Phi_d(\vq)$ by $i\vq$ and then inverse Fourier transforming yields the modes of $\grad \Phi_b(\vx)$ and $\grad \Phi_d(\vx)$. We also inverse Fourier transform the other two fields, $\Phi_a$ and $\Phi_c$ and, now in position space, form the products $\vec{V}_{ab}(\vx) =  \Phi_a(\vx) \grad\Phi_b(\vx) $ and $\vec{V}_{cd}(\vx) = \Phi_c(\vx) \grad \Phi_d(\vx)$. Next, we Fourier transform both of these and, from $\vec{V}_{cd}(\vq)$, solve for the modes of $\vec A_{cd}(\vq)$ using Eq.~\eqref{eq:amodes}. The unbinned vector POP spectrum estimator is
\begin{align}
    P_{\mathrm{vector}}(k) =  \frac{1}{V_{\mathrm{box}}N_{\mathrm{vector}}(k)} \sum_{|\vq|=k} \vec V_{ab}(\vq) \cdot\vec A^*_{cd}(\vq) \, ,
\end{align}
where the sum is over all modes with equal wave vector magnitudes on the grid.

The normalization is computed by initializing four grids of modes corresponding to the filter functions: $f_a(\vq)$, $f_b(\vq)$, $f_c(\vq)$, and $f_d(\vq)$. From here, obtaining the normalization corresponding to Eq.~\eqref{eq:vpvnorm} is somewhat complicated. When constructing the power spectrum, we get one triple product from the estimator and another from the parity-odd trispectrum. Overall, the power spectrum involves the square of the triple product. To get a squared triple product in the normalization factor, we need to compute the Hessian of the field $f_a(\vx)$ and construct the composite tensor field with components
\begin{align} \label{eq:hij}
    F^{ij}_{ab}(\vx) = f_a(\vx)\nabla^{i} \nabla^{j} f_b(\vx) \, .
\end{align}
After Fourier transforming the Hessian of the filter function fields, we take the curl of one of the two components,
\begin{align}
    J_{ab}^{ij}(\vq) \equiv \epsilon^{imn} i k^m F^{nj}_{ab}(\vq) \, ,
\end{align}
where repeated indices imply summation. Following the same steps for $f_c(\vx)$ and $f_d(\vx)$, the desired normalization is
\begin{align}
    N_{\mathrm{vector}}(k) = -\sum_{|\vq|=k} J_{ab}^{ij}(\vq) \Big[J_{cd}^{ij}(\vq)\Big]^{\dag}  \, .
\end{align}
The dagger $\dagger$ denotes the Hermitian conjugate of the matrix. We then bin the estimator in wave number bins of width $\Delta k$,
\begin{align}
    P_{\mathrm{vector}}(k|\Delta k) = \frac{1}{N_{\mathrm{vector}}(k|\Delta k)} \sum_{q \in k} P_{\mathrm{vector}}(q) N_{\mathrm{vector}}(q) \, ,
\end{align}
where
\begin{align}
    N_{\mathrm{vector}}(k|\Delta k) = \sum_{q \in k} N_{\mathrm{vector}}(q) \, .
\end{align}
Here, the sum over $q \in k$ means summing over all $q \in [k-\Delta k/2, k+\Delta k/2)$.

The computation of the scalar POP spectrum follows similar steps. In this case, we compute the gradients $\grad \Phi_a(\vx)$, $\grad \Phi_b(\vx)$ and $\grad \Phi_c(\vx)$. We use the latter two to form $\vec B_{bc}(\vx)$ according to Eq.~\eqref{eq:pvecb}. We Fourier transform this and then multiply by the filtering function modes $f_d(\vq)$, resulting in the modes of $\vec B_{bcd}(\vq)$. We then inverse Fourier transform this, and its dot product with $\grad \Phi_a(\vx)$ yields $\Psi_{abcd}(\vx)$. The unbinned scalar POP spectrum is
\begin{align}
    P_{\mathrm{scalar}}(k) =  \frac{1}{V_{\mathrm{box}}N_{\mathrm{scalar}}(k)} \sum_{|\vq|=k} \Phi(\vq)\left[\Psi_{abcd}(\vq)\right]^{*} \, .
\end{align}

Constructing the normalization is again nontrivial due to the squared triple product factor. In this case, we require the Hessians of the field $f_a(\vx)$, $f_b(\vx)$, and $f_c(\vx)$. From the latter two, we construct the tensor field $F^{ij}_{bc}(\vx)$ as in Eq.~\eqref{eq:hij}. By Fourier transforming all components and multiplying the modes by the filter function $f_d(\vq)$, we obtain $F^{ij}_{bcd}(\vq) \equiv f_d(\vq) F^{ij}_{bc}(\vq)$. Inverse Fourier transforming this field and multiplying by the Hessian of $f_a(\vx)$ gives
\begin{align}
    G^{ij}_{abcd}(\vx) = \epsilon^{imn} \epsilon^{jpq} F^{mp}_{bcd}(\vx) \nabla^n \nabla^q f_a(\vx)  \, .
\end{align}
Fourier transforming this, the scalar POP spectrum normalization is
\begin{align}
    N_{\mathrm{scalar}}(k) = -\sum_{|\vq|=k} k^i k^j G^{ij}_{abcd}(\vq) \, .
\end{align}
The binned estimator is
\begin{align}
    P_{\mathrm{scalar}}(k|\Delta k) = \frac{1}{N_{\mathrm{scalar}}(k|\Delta k)} \sum_{q \in k} P_{\mathrm{scalar}}(q) N_{\mathrm{scalar}}(q) \, ,
\end{align}
where
\begin{align}
    N_{\mathrm{scalar}}(k|\Delta k) = \sum_{q \in k} N_{\mathrm{scalar}}(q) \, .
\end{align}

If we choose filter functions that only select wave vector magnitude shells of width $\Delta k$, then both POP spectra coincide and equal the binned trispectrum estimator in Eq.~\eqref{eq:tbin}. This fact illustrates the computational advantage of our POP spectra. Evaluating all the bins of one POP spectrum is computationally equivalent to constructing a one-dimensional slice of the full trispectrum. The tradeoff is that we have significantly compressed the information of the full trispectrum. However, the presence of two estimators, each carrying complementary information from the parity-odd trispectrum, partially compensates for this reduction in information. We also have the freedom to choose the filter functions, which can be optimized. The extent of information loss could be further mitigated by considering extensions and generalizations of the POP spectra, which we explore in \S\ref{sec:extension}.

The algorithms that we presented above for constructing these estimators are efficient since the most costly operations are the Discrete Fourier Transforms (DFTs), which scale as $N\log N$, with $N$ the number of grid points. All other operations are point-wise multiplications, which scale as $N$. Ignoring the normalization factors, the vector POP spectrum requires 15 and the scalar POP spectrum requires 17 3D DFTs. Constructing the estimators without the normalization factor takes about 6 seconds for the vector and 8 seconds for the scalar POP spectrum running on a single node with 128 cores. The normalization factors are more expensive to compute, but these only need to be computed once and can be saved and reused since they are not data-dependent. 

\section{Validation of the Estimators}
\label{sec:valid}

\subsection{Random Parity-Violating Realizations}
\label{subsec:rand}

\begin{figure}
    \centering
    \includegraphics[width = 0.9\linewidth]{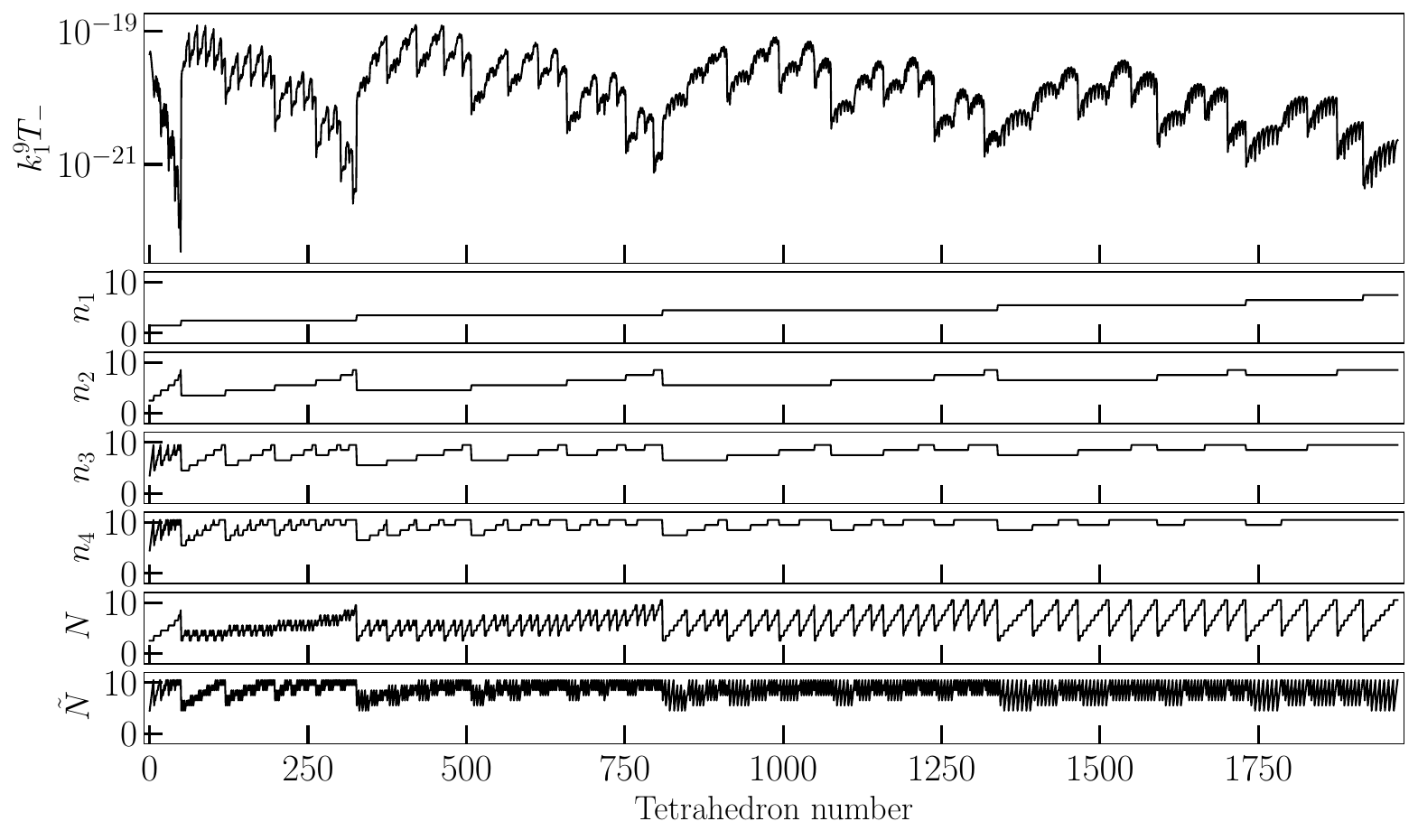}
    \caption{Here we show the trispectrum in the thin bin limit (top panel) and the binning scheme (lower six panels). The shape of the trispectrum template from Eq.~\eqref{eq:template} appears in the top panel. The bottom six panels indicate the binning scheme with $n_i = k_i / k_{\rm F}$, $N = K / k_{\rm F}$, and $\tilde{N} = \tilde{K} / k_{\rm F}$. Here, $k_{\rm F}$ is the fundamental mode of a box interpreted as having length 4~${\rm Gpc}~h^{-1}$. We impose $k_1 < k_2 < k_3 < k_4$. We have used 10 $k$-bins of width $k_{\rm F}$.}
    \label{fig:tspec}
\end{figure} 

To validate our POP estimators, we generate a set of random primordial curvature fields that include a specific parity-violating trispectrum. We start by generating the modes of a Gaussian random field primordial curvature perturbation $\zeta_{\mathrm{G}}(\vk)$,
\begin{align}
    \zeta_{\mathrm{G}}(\vk) = \sqrt{\frac{V_{\mathrm{Box}}\, \pi^2 A_\mathrm{s}}{ k^3}
    \left(\frac{k}{k_\mathrm{p}}\right)^{n_\mathrm{s}-1}} \, \big[N_1(\vk) + i N_2(\vk)\big],
\end{align}
where $N_1$ and $N_2$ are drawn from a standard normal distribution, $A_\mathrm{s}$ is the amplitude of the primordial power spectrum, $n_\mathrm{s}$ is the spectral tilt, and $k_\mathrm{p}=0.05~\mathrm{Mpc}^{-1}$ is the pivot scale. We then transform this Gaussian random field into a non-Gaussian field in real space as \citep{Coulton:2023oug}
\begin{align} \label{eq:phi}
    \zeta(\vx) = \zeta_{\mathrm{G}}(\vx) + g_{-} \grad  \zeta_{\mathrm{G}}^{[\alpha]}(\vx)  \cdot \left[ \grad  \zeta_{\mathrm{G}}^{[\beta]}(\vx)  \times \grad  \zeta_{\mathrm{G}}^{[\gamma]}(\vx) \right] \, .
\end{align}
The coefficient $g_{-}$ controls the amplitude of the primordial parity-odd trispectrum, and the modes of the fields in the triple product are
\begin{align}
    \label{eq:phinl}
    \zeta_{\mathrm{G}}^{[\alpha]}(\vk) = k^\alpha \zeta_{\mathrm{G}}(\vk) \, .
\end{align}
The leading order imaginary part of the trispectrum for this template is given by
\begin{align} \label{eq:template}
T_{-}(\vk_1, \vk_2, \vk_3) = g_{-}\, \vk_1 \cdot (\vk_2 \times \vk_3) \left(2 \pi^2 A_\mathrm{s}\right)^3 \left(k_1^{\alpha-4 + n_\mathrm{s}} k_2^{\beta-4 + n_\mathrm{s}} k_3^{\gamma-4 + n_\mathrm{s}} k_4^{0} \mp \mathrm{23\ signed\ permutations}\right) \, .
\end{align}
The 24 terms in the parenthesis are the permutations of $\{k_1, k_2, k_3, k_4\}$. Even permutations (with an even number of transpositions) get a positive sign, and odd permutations get a negative sign in the sum. We have included the $k_4^0$ factor to emphasize that every term has one of the four wave vector magnitudes raised to the exponent zero. 

The trispectrum is scale invariant if $\alpha+\beta+\gamma=-3$. To preserve the large-scale power spectrum we also restrict these exponents to be less than or equal to zero. We choose $\alpha=-2$, $\beta=-1$, and $\gamma=0$, which is the same template used in \citet{Coulton:2023oug}. For simplicity, we choose $A_\mathrm{s} = 2\times 10^{-9}$, $n_\mathrm{s} = 1$, and $g_{-}=\pm10^6$. In Fig.~\ref{fig:tspec}, we display the shape of this trispectrum.

Choosing a binning scheme where $k_1 < k_2 < k_3 < k_4$, the trispectrum shape function is dominated by the terms
\begin{align}
    \label{eq:taumax}
    \tau_-(k_1,k_2,k_3,k_4) \simeq \frac{g_-(2\pi^2 A_s)^3}{k_1^5 k_2^4} \Big( \frac{1}{k_3^3} - \frac{1}{k_4^3} \Big) \, .
\end{align}
This shape peaks when the wave vectors $\vk_1$, $\vk_2$, and $\vk_3$ are parallel so that $k_4 = k_1 + k_2 + k_3$. The diagonals for this configuration are $K=k_1+k_2$ and $\tilde{K}=k_4-k_1$. However, this configuration is insensitive to parity, since the triple product vanishes when any two wave vectors are colinear. The trispectrum peaks on a tetrahedron shape that deviates from colinearity with diagonals $K<k_1+k_2$ and $\tilde{K}>k_4-k_1$ that maximize the product of the right-hand side of Eq.~\eqref{eq:taumax} and the triple product $\vk_1 \cdot (\vk_2\times\vk_3)$. This peak analysis is true for any scale-invariant trispectrum of the form in Eq.~\eqref{eq:template}, although the particular diagonals that maximize the trispectrum depend on the values of the template exponents. These parity-odd trispectrum templates diverge in the soft limit $k_1\rightarrow0$.

\begin{figure}
    \centering
    \includegraphics[width = 0.8\linewidth]{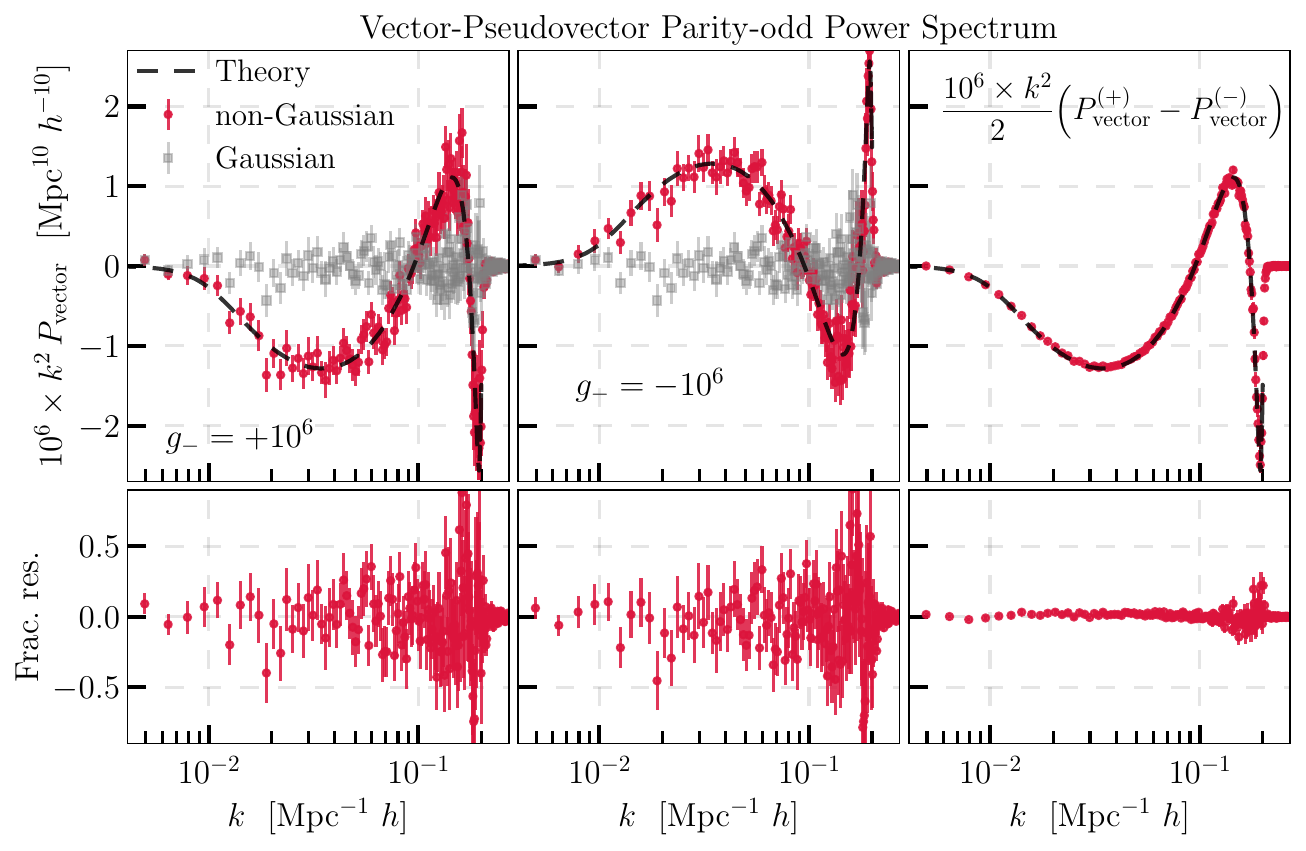}
    \caption{The vector POP spectrum of the primordial potential defined in Eq.~\eqref{eq:phi}. In the left panels, the amplitude of the primordial trispectrum is $g_-=10^6$, while in the middle panels $g_-=-10^6$. We used the same underlying Gaussian realizations for both signs of $g_-$. The right panels show the variance-suppressed estimator, taking the difference between results with positive and negative $g_-$. The black dashed curve is the expected signal, computed from Eqs.~(\ref{eq:vpvth}--\ref{eq:vecnorm}). The bottom row shows fractional residuals of the simulated data with respect to the semi-analytical calculation. Gray data points show the estimator on a Gaussian random field. These scatter around zero and are consistent with a vanishing parity-violating signal. All data points have error bars obtained by bootstrap averaging over the 64 simulations, resampling with replacement $10^{5}$ times. In the right-most panel, the sample variance cancellation is not perfect. This is because the POP spectrum involves four weakly non-Gaussian fields.}
    \label{fig:vpv}
\end{figure}

Since we are dealing with a scale-invariant primordial potential, the box length is arbitrary, but we interpret it as $L_{\mathrm{Box}}=4~\mathrm{Gpc}~h^{-1}$ so that the scales involved are relevant for CMB and LSS. We generate 64 random pairs of these non-Gaussian potential fields on a grid of size $N_{\mathrm{grid}} = 512^3$. The pairs have the same underlying Gaussian realization but opposite signs for their nonlinear terms. Subtracting the POP spectra measured on these pairs suppresses cosmic variance from the purely Gaussian part of the fields. As can be seen in Fig.~\ref{fig:vpv}, the cosmic variance is not perfectly cancelled. This is because the POP spectrum is a compression of the four-point correlator of a weakly non-Gaussian field.

There is a considerable amount of freedom in our estimators since we can choose any set of filtering functions. For the vector POP spectrum, $f_a(k)$ must be distinct from $f_b(k)$, and $f_c(k)$ must be distinct from $f_d(k)$, otherwise $P_{\rm vector}(k)$ vanishes identically. For the scalar POP spectrum, $f_a(k)$, $f_b(k)$, and $f_c(k)$ must all be distinct for $P_{\rm scalar}(k)$ not to vanish. Here, we choose
\begin{align}
	f_a(k) = k^2 \Theta(k - k_{\mathrm{min}})  \Theta(k_{\mathrm{max}} - k) \, , \\
	f_c(k) = k^{-2} \Theta(k - k_{\mathrm{min}})  \Theta(k_{\mathrm{max}} - k) \, , \\
	f_b(k) = f_d(k) =  \Theta(k - k_{\mathrm{min}})  \Theta(k_{\mathrm{max}} - k)  \, .
\end{align} 
Each filtering function involves the same scale cuts, $k_{\mathrm{min}} = 5\times10^{-3}~\mathrm{Mpc}^{-1}~h$ and $k_{\mathrm{max}} = 2\times10^{-1}~\mathrm{Mpc}^{-1}~h$, which selects a thick spherical shell of modes. As long as the fundamental mode in the box is less than $k_{\mathrm{min}}$, and the Nyquist mode is greater than $k_{\mathrm{max}}$, the box contains the full range of modes and the estimator should be independent of the box size and resolution. However, the discreteness of the grid is noticeable on large scales if $k_{\mathrm{min}}$ is very close to the fundamental mode.

The autocorrelation of the nonlinear term in Eq. \eqref{eq:phi} will introduce corrections to the power spectrum. These corrections are of the order $\sim g_{-}^2 A_\mathrm{s}^3$ and are subdominant compared to the Gaussian power spectrum for our template at the scales we consider. The non-Gaussian corrections to the power spectrum grow at small scales and increase the power of the Nyquist mode, $k=0.4~\mathrm{Mpc}^{-1}~h$, in our simulated boxes by about 1\%. After imposing the scale cuts in our POP spectra the corrections to the power spectrum are far below percent level for the modes we analyze. These corrections to the power spectrum can be removed by further rescaling the non-Gaussian field \citep{Coulton:2023oug}. However, corrections to the power spectrum could be a physical effect. The (inflationary) mechanisms that generate the primordial trispectrum can also affect the shape of the power spectrum. This template does not introduce a primordial bispectrum because all contributions would involve an odd number of Gaussian fields, resulting in a vanishing expectation value.

We have omitted many complications of observational survey data in this analysis. Cosmological surveys target observables that are biased tracers of the underlying matter field, which has undergone nonlinear evolution. Accurate modelling will require the linear matter transfer function, the bias expansion, and possibly the non-linearity of gravitational clustering depending on the scales considered. Surveys observe objects in redshift space where isotropy does not hold. Observational analyses must also model the survey mask, selection function, and shot noise. We will investigate the impact of these aspects of observational data on the POP spectra in future work.

\subsection{Comparison with Analytical Computation}
\label{sec:res}

\begin{figure}
    \centering
    \includegraphics[width = 0.8\linewidth]{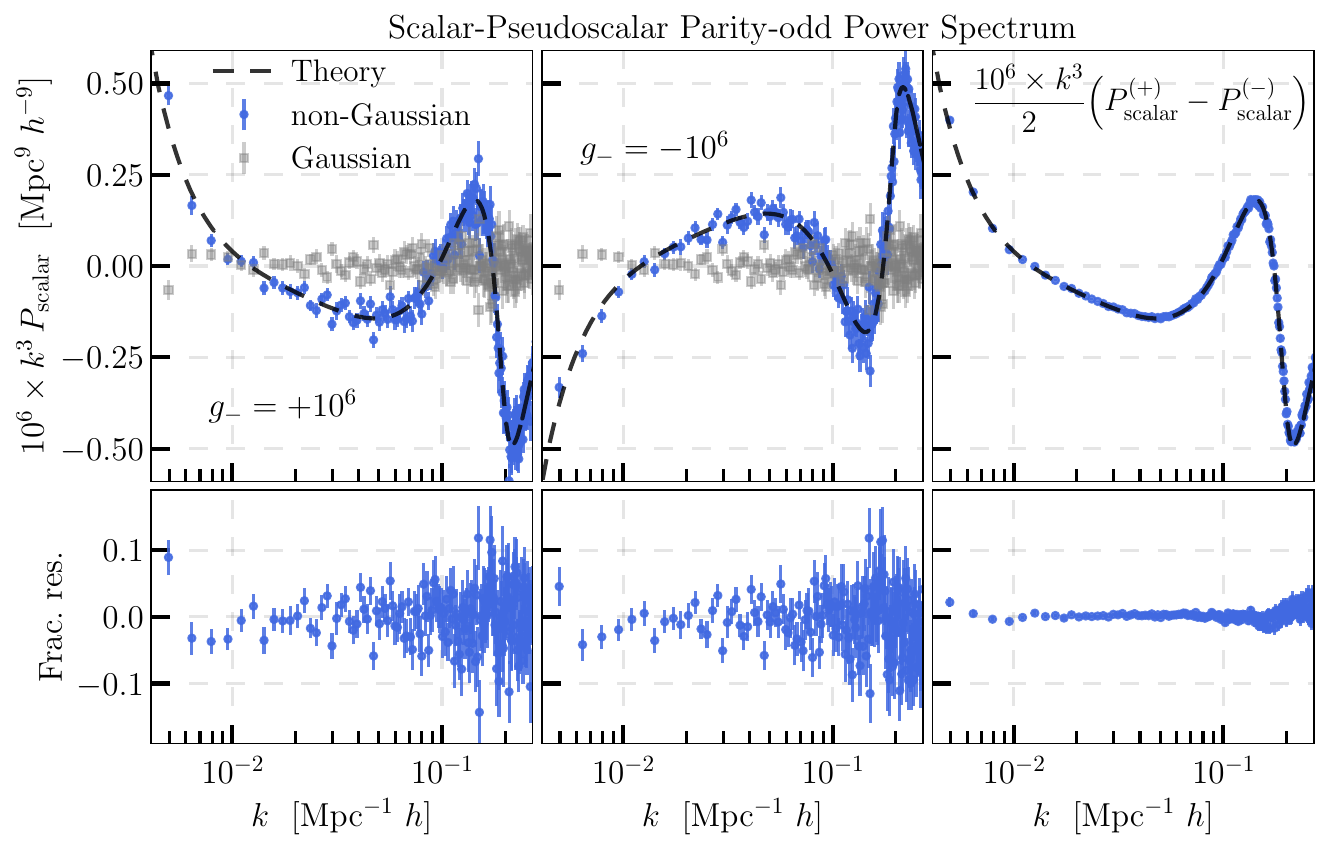}
    \caption{Same as Fig.~\ref{fig:vpv} but for the scalar POP spectrum. In this case, the signal grows in the limit $k\rightarrow0$, showing sensitivity to the soft limit of the trispectrum as one of the tetrahedron wave vector side lengths becomes small. In contrast, Fig.~\ref{fig:vpv} demonstrates decreasing sensitivity in the diagonal soft limit of the vector POP spectrum because the specific trispectrum template that we simulate does not peak in this limit. The behavior of the residual in the right-most panel is again (as in Fig.~\ref{fig:vpv}) due to the additional sample variance contributed by the weakly non-Gaussian nature of the fields.}
    \label{fig:sps}
\end{figure}

In Fig.~\ref{fig:vpv}, we display the vector POP spectrum compared with the theoretical expectation obtained by numerically evaluating the integrals in Eqs.~(\ref{eq:vpvth}--\ref{eq:vecnorm}). The left panel shows the results for 64 simulations with $g_-=+10^6$. The middle panel shows results for the same underlying Gaussian random fields but with $g_-=-10^6$. We significantly reduce cosmic variance by computing the difference between these two estimators, as illustrated in the right panel of Fig.~\ref{fig:vpv}. We estimated the error bars by bootstrap averaging over the estimators constructed for the 64 realizations, resampling $10^5$ times with replacement.

We obtained the semi-analytic result in Fig.~\ref{fig:vpv} by evaluating Eq.~\eqref{eq:vpvth} on a discretized lattice for the radial integrals over all $q_i$ and $k=|\vq_1 + \vq_2|$. Specifically, for a given bin, we construct a three-dimensional grid of values for $\{q_1, q_2, k\}$, with $q_1$ and $q_2$ spaced uniformly between $k_{\mathrm{min}}$ and $k_{\mathrm{max}}$ and $k$ spanning the width of the power spectrum bin. We then mask out regions that violate the triangle inequalities. Next, we construct a second grid for $\{q_3, q_4, k\}$, masking the non-triangular regions. Finally, we integrate Eq.~\eqref{eq:vpvth} term by term. For each permutation term in the trispectrum of Eq.~\eqref{eq:template}, we construct the integrand for $q_1$ and $q_2$, and compute the discretized integral using Romberg integration with $2^8+1$ sample points. We do the same for $q_3$ and $q_4$, except we include explicit factors of $k$ appearing in Eq.~\eqref{eq:template} in this grid. We then multiply the results from the two grids. Finally, we compute the discretized integral of the diagonal $k$ using Romberg integration with $2^7+1$ sample points. The procedure for integrating Eq.~\eqref{eq:vecnorm} is the same, omitting the factors from the primordial trispectrum template shape.  We parallelize this calculation by computing multiple bins simultaneously. From Fig.~\ref{fig:vpv}, we see excellent agreement with the semi-analytic calculation of the expected signal. This agreement demonstrates that we accurately recover the expected signal from the trispectrum shape injected into the data.

In Fig.~\ref{fig:sps}, we display similar results for the scalar POP spectrum. The theoretical calculation of Eqs.~(\ref{eq:spsth}--\ref{eq:scalarnorm}) differs from the vector case in that $k$ in the triangle $\{k, q_2, Q\}$ is integrated over the estimator's bin width spanning the range between $k_{\mathrm{min}}$ and $k_{\mathrm{max}}$ and the filter function $f_d(Q)$ applies to the diagonal $Q$. Again, we find excellent agreement between our simulated data and the expected shape of $P_{\rm scalar}(k)$. The bias at low $k$ is due to the discreteness of the grid, compared with the continuous-limit integrals computed in the semi-analytic prediction. This bias is not noticeable for the vector estimator in Fig.~\ref{fig:vpv} due to the signal vanishing at low $k$.

As a final remark note that, unlike the vector POP spectrum, the scalar POP spectrum increases as $k$ approaches zero. This occurs because the argument $k$ of the scalar POP spectrum corresponds to the side length of the tetrahedron. Our specific trispectrum template, Eq.~\eqref{eq:template}, diverges as one side length goes to zero, so the scalar POP spectrum also diverges in this limit. If we chose a template that peaks in the soft limit where a diagonal approaches zero, the vector POP spectrum would increase as its argument approaches zero. We could also construct a template that diverges in the soft side length and soft diagonal limits. Then the vector and scalar POP spectra would increase as their arguments go to zero. This demonstrates that the $k\rightarrow 0$ limits of our POP spectra encode crucial and complementary information about the soft limits of the trispectrum.

\begin{figure}
    \centering
    \includegraphics[width = 0.8\linewidth]{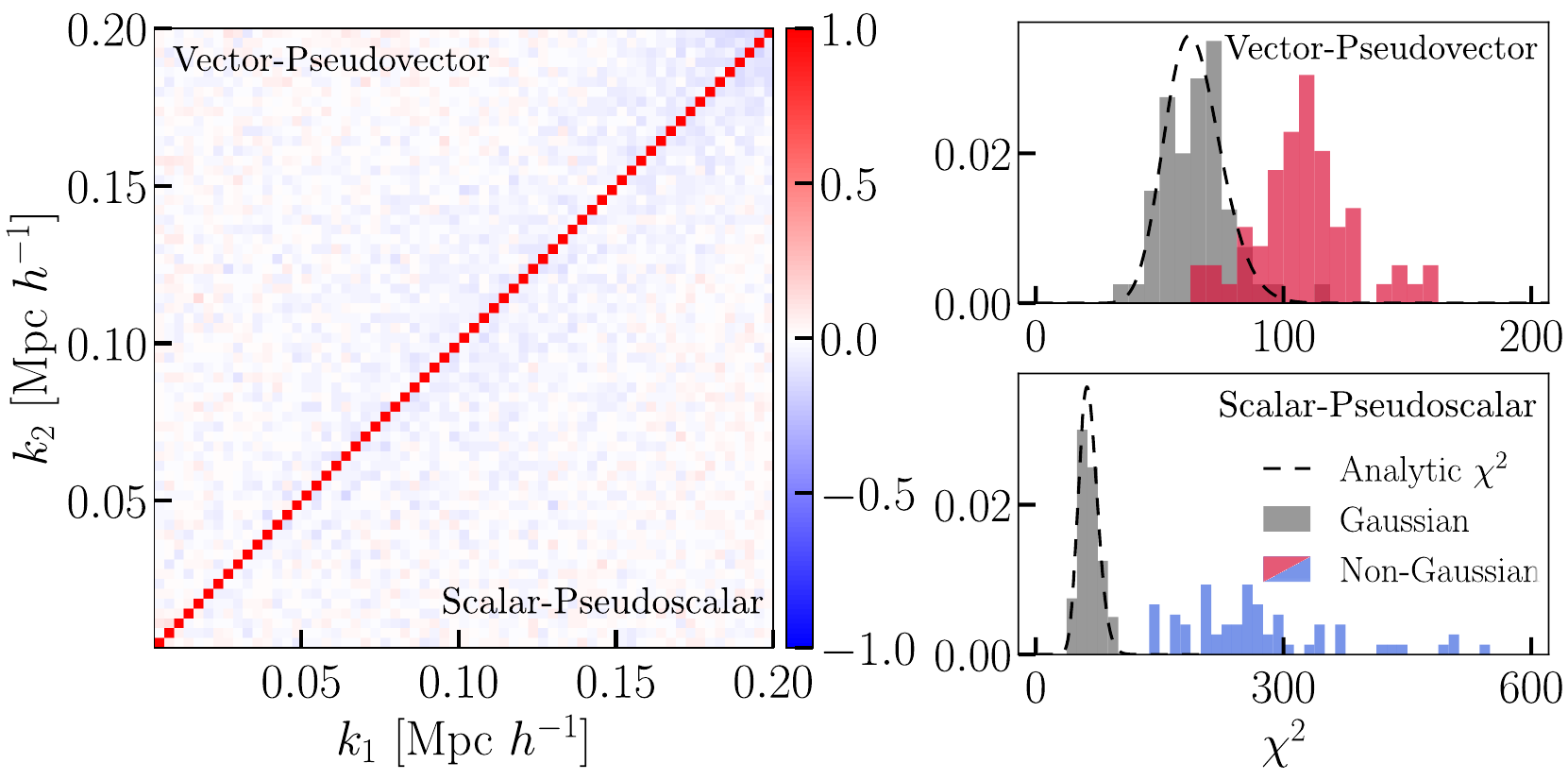}
    \caption{On the left panel, we display the correlation matrix, Eq.~\eqref{eq:corr}, for the vector-pseudovector (upper triangle) and scalar-pseudoscalar (lower triangle) estimators. The correlation matrix is dominated by the diagonal, with off-diagonal contributions less than 10\%. On the right panels, we display histograms of the $\chi^2$ values from 64 random non-Gaussian fields compared with the distribution from Gaussian realizations in gray. The peak of the dashed, analytical $\chi^2$ curve coincides with the number of bins analyzed.}
    \label{fig:corr}
\end{figure}

\subsection{Correlation Matrix and \texorpdfstring{$\chi^2$}{x\^2} Test}

The POP spectra integrate over many configurations to obtain a power spectrum from a trispectrum. For this reason, we may worry that the covariance matrices of the POP spectra have complicated structures with large off-diagonal contributions. To test this, we generate 1024 realizations of random Gaussian fields, construct the vector and scalar POP spectra estimators, and compute the covariance matrix for each,
\begin{align}
    C(k_i, k_j) = \langle P(k_i) P(k_j) \rangle - \langle P(k_i) \rangle \langle P(k_j) \rangle \, ,
    \label{eq:corr}
\end{align}
where $k_i$ indicates the $i^{\mathrm{th}}$ wave number bin, and $P$ can be either $P_{\mathrm{vector}}$(k) or $P_{\mathrm{scalar}}$(k). In Fig.~\ref{fig:corr} we display the correlation matrices,
\begin{align}
R(k_i, k_j) = \frac{C(k_i, k_j)}{\sqrt{C(k_i,k_i) C(k_j,k_j)}} \ ,
\end{align}
for the vector POP spectrum in the upper-right triangle and the scalar POP spectrum in the lower-left triangle. We find the correlation matrix is dominated by its diagonal for both estimators, with off-diagonal contributions less than $10\%$.

We assess the sensitivity of our compressed estimators by computing the $\chi^2$ from the inverse of the covariance matrix,
\begin{align}
\chi^2 = \sum_{ij} P(k_i) C^{-1}(k_i, k_j) P(k_j) \, .
\end{align}
We compute the $\chi^2$ values for all 64 of our parity-violating non-Gaussian fields with positive amplitude for the primordial trispectrum and their underlying Gaussian fields. Fig.~\ref{fig:corr} displays histograms of the $\chi^2$ values. The scalar POP spectrum exhibits greater sensitivity for detecting primordial parity violation than the vector POP spectrum. We can understand this from the form of primordial non-Gaussianity in Eq.\eqref{eq:phi} that leads to the shape of the trispectrum in Eq.~\eqref{eq:template}. This shape peaks as $k_1$, the argument of the scalar POP spectrum, goes to zero. We see this in the low-$k$ behaviour in Fig.~\ref{fig:sps}.

The sensitivity depends on the choice of filtering kernels and the shape of the primordial signal. Different parity-odd trispectrum templates will generally yield different $\chi^2$ distributions with more or less sensitivity. In principle, we can find the optimal filter functions to maximize the signal-to-noise ratio of the POP power spectra for a given shape of the underlying primordial trispectrum, as was done for the CMB lensing estimator \citep{Hu:2001fa}, but we do not explore that in this work. Since the vector and scalar POP spectra contain complementary information, it would be best to do a joint analysis of both simultaneously. This joint analysis requires the joint covariance, which will have off-diagonal contributions between the vector and scalar POP spectra bins. We leave the analysis of the joint covariance matrix to future work.

\section{Extensions and Generalizations}
\label{sec:extension}
The examples of parity-odd power spectra explored above are the simplest examples of a more general class of constructions. In this section, we outline some of these generalizations.

\subsection{Connection to the Polarization Basis}

Before exploring some generalizations of our POP spectra, we first write down an alternative and equivalent definition of the vector POP spectrum based on polarization vectors, which will be useful later. As described in Sec.~\ref{sec:ppvt}, parity violation is closely related to right- and left-handed helicities. Here we will demonstrate the relationship between our vector POP spectrum and these helicities. To do so, let us take two vector fields $\vec{V}_{ab}$ and $\vec{V}_{cd}$. From these, we can cross-correlate their Cartesian components as
\begin{align}
    \langle V^i_{ab}(\vk) V^j_{cd}(\vk') \rangle = (2\pi)^3 \delta_{\rm D}^{(3)}(\vk+\vk') P^{ij}(k) \, .
\end{align}
Assuming isotropy, we can decompose this power spectrum matrix into its trace, its traceless symmetric,  and its antisymmetric parts,
\begin{align}
P^{ij}(k) = \frac{1}{3}\delta^{ij} P_{\parallel}(k) + \Big(\frac{k^ik^j}{k^2} - \frac{1}{3}\delta^{ij}\Big) P_{\perp}(k) + i \epsilon^{ijk} \frac{k^k}{k} P_{-}(k) \, .
\end{align}
Assuming parity invariance, only the first two terms are present. These are the longitudinal and transverse parts of the parity-even power spectrum matrix. The final term is the parity-odd part of the power spectrum matrix.

One can choose left-/right-handed polarization vectors and form a complete basis $\{\vec{e}_\parallel(\vk), \vec{e}_\mathrm{R}(\vk), \vec{e}_\mathrm{L}(\vk)\}$ in Fourier space,
\begin{align}
\vec{V}_{ab}(\vec{k}) = V_{ab,\parallel}(\vec{k}) \vec{e}_\parallel(\vk) + 
 V_{ab,\mathrm{R}}(\vec{k}) \vec{e}_\mathrm{R}(\vk) + 
V_{ab,\mathrm{L}}(\vec{k}) \vec{e}_\mathrm{L}(\vk) \, .
\end{align}
The transverse polarization vectors are defined as eigenvectors of the curl operator,
\begin{align}
\label{eq:epsilon}
    i\vk \times \vec{e}_\mathrm{R/L}(\vk) \equiv \pm k \, \vec{e}_\mathrm{R/L}(\vk) \, ,
\end{align}
and $\vec{e}_\parallel(\vk) \equiv i\vk/k$. These quantities transform into one another after a parity transformation, as $\vec{e}_\mathrm{R/L}(-\vk)=\pm i\vec{e}_\mathrm{L/R}(\vk)$. Therefore, one can define the eigenvectors of the parity operator as
\begin{align}
\vec{e}_\pm(\vk) = \frac{1}{\sqrt{2}}\big(\vec{e}_\mathrm{R}(\vk) \pm i \vec{e}_\mathrm{L}(\vk)\big) \, .    
\end{align}
Using this alternative basis, the vector field is decomposed as
\begin{align}
\vec{V}_{ab}(\vec{k}) = V_{ab,\parallel}(\vec{k}) \vec{e}_\parallel(\vk) + 
 V_{ab,-}(\vec{k}) \vec{e}_+(\vk) + 
V_{ab,+}(\vec{k}) \vec{e}_-(\vk) \, .
\end{align}
The $\pm$ labels of the coefficients are the opposite of the basis labels because the vector must be parity odd, so the coefficients transform with the opposite sign compared to the transverse basis vectors. We can then project the power spectrum matrix onto the polarization vectors, 
\begin{align}
    P_{\rm RR/LL}(k) = P^{ij}(k) e^{i}_{\rm R/L}(\vk) e^j_{\rm R/L}(-\vk) \, ,
\end{align}
or 
\begin{align}
    P_{+-}(k) = P^{ij}(k) e^i_{+}(\vec{k}) e^j_{-}(-\vec{k}) \, .
\end{align}
It is straightforward to show that
\begin{align}
P_{-}(k) & = \ -\frac{k}{2} P_{\rm vector}(k) \, \\
         & = \ \frac{1}{2}\Big( P_{\mathrm{RR}}(k) - P_{\mathrm{LL}}(k) \Big) \, \\
         & = \ P_{+-}(k) \, ,
\end{align}
where $P_{\rm vector}(k)$ is the vector POP spectrum.

\subsection{Higher-Order Derivatives: Angular Dependence}

The vector and scalar POP spectra that we have considered thus far are minimal in the sense that they involve the minimum number of spatial derivatives required to form the parity-odd triple product. We are free to add any number of spatial derivatives to these constructions, forming higher-rank tensor POP spectra. For example, we consider adding an additional derivative to the vector and pseudovector fields from the vector POP spectrum, as
\begin{align}
    H_{1{\rm ab}}^{ij}(\vx) = \nabla^i \Phi_{\rm a}(\vx)  \nabla^j \Phi_{\rm b}(\vx) \, , \\
    H_{2{\rm cd}}^{ij}(\vx) = \epsilon^{ikl} \nabla^j \nabla^k \Phi_{\rm c}(\vx)  \nabla^l \Phi_{\rm d}(\vx) \, .
\end{align}
The POP spectrum formed by contracting the indices of these two fields $\langle H_{1{\rm ab}}^{ij}(\vk) H_{2{\rm cd}}^{ij}(\vk')  \rangle$ has a similar form to the vector POP spectrum in Eq.~\eqref{eq:vpv}, the difference being that the integrand would now contain an additional factor of $\vq_1\cdot\vq_3$. Therefore, this POP tensor spectrum probes a different form of angular dependence in the parity-odd trispectrum.
 
We can also define higher-rank tensor POP spectra by contracting $\langle H_{1\rm{ab}}^{li}(\vk) H_{2\rm{cd}}^{lj}(\vk')  \rangle$. We can then project this onto the basis tensors
\begin{align}
e^{ij}_{\rm R/L}(\vk) = e^i_{\rm R/L}(\vk)  e^j_{\rm R/L}(\vk) \, ,
\end{align}
where $e^i_{\rm R/L}(\vk)$ are components of the polarization vectors, defined in Eq.~\eqref{eq:epsilon}. This decomposition gives the right-handed and left-handed components of the transverse tensor spectra, and the difference between these is a POP spectrum. This kind of construction can be easily generalized to higher-derivative and higher-rank tensor POP spectra. Some specific examples of these constructions can be found in \citet{Jeong:2012df} and \citet{Shim:2024tue}. Similar constructions for the scalar POP spectra are also possible. 

\subsection{Higher-Order Constructions: Parity-Odd Bispectra and Beyond}

The scalar bispectrum is defined by the three-point correlation function in Fourier space through
\begin{align}
    \langle \Phi(\vec{k}_1) \Phi(\vec{k}_2) \Phi(\vec{k}_3) \rangle = (2\pi)^3 \delta_{\mathrm{D}}^{(3)}(\vk_1 + \vk_2 + \vk_3) B(k_1, k_2, k_3) \, .
\end{align}
The Dirac delta function enforces statistical translational invariance and makes the bispectrum a function over two-dimensional, closed triangles. Since in three dimensions, we can rotate a parity-transformed two-dimensional triangle back to its initial configuration, the bispectrum is insensitive to parity under statistical isotropy.

We construct the parity-odd bispectrum that compresses the parity-odd four-point function in a similar way to the vector POP spectra,
\begin{align}
    \label{eq:pob}
    \langle \vec{A}_{ab}(\vec{k}_1) \cdot i \vec{k_2} \Phi(\vec{k}_2) \Phi(\vec{k}_3) \rangle = (2\pi)^3 \delta_{\mathrm{D}}^{(3)}(\vk_1 + \vk_2 + \vk_3) B_{\mathrm{vector}}(k_1, k_2, k_3) \, .
\end{align} 
In this case, we have used the gradient of the scalar rather than the composite vector field. With the composite vector field, this construction would involve five scalar fields corresponding to a compression of the parity-odd five-point function. Similarly, the pseudoscalar requires a product of three scalar fields for its constructions, so the parity-odd bispectrum of one composite pseudoscalar with two scalar fields compresses the parity-odd five-point statistics. Additionally, we can correlate the vector-pseudovector inner product with two pseudoscalar fields or correlate three pseudoscalar fields. These constructions are of order six in the underlying scalar field, compressing the parity-odd six-point function. Thus, Eq. \eqref{eq:pob} is the only form for a parity-odd compression of the trispectrum to a bispectrum.

\section{Conclusion}
\label{sec:conclude}
Exploring parity violation on large cosmological scales will provide crucial insights into the fundamental symmetries governing the origins and evolution of the Universe. The search for a parity-odd signal within galaxy surveys presents a key challenge in observational cosmology, especially in light of recent hints of a parity-odd signal in the four-point correlation function of galaxies \citep{Hou:2022wfj, Philcox:2022hkh}, both of which used the isotropic basis functions of \cite{Cahn:2020axu}, the method proposed in \cite{Cahn:2021ltpa}, and the covariance template of \cite{Hou:2021ncj}. Direct measurements of four-point correlations pose several technical challenges related to the high dimensionality of the data vector and the resulting size of the covariance matrix, making it difficult to interpret the signal and assess its statistical significance. Motivated by these challenges, we developed a novel set of statistics for parity violation.

The new observables are two-point correlators calculated between composite fields derived via nonlinear transformations of the original scalar field. These constructions compress the trispectrum information into a power spectrum, resulting in a significant simplification compared to the full four-point correlation function. The advantages of this approach are two-fold. First, the computation of a two-point function is faster because of the significant dimensionality reduction of the data vector; our computation on a $512^3$ grid takes fewer than 10 seconds for each spectrum on 128 cores. This reduction facilitates a more efficient and interpretable analysis of the underlying parity-odd signal and its covariance. Second, formulating the estimators in Fourier space enables efficient characterization of the scale dependence of the signal and its soft limits.

We defined two sets of estimators: one based on correlating a vector and a pseudovector field and another correlating a scalar and a pseudoscalar field, both derived from the original scalar field. These two estimators are distinct compressions of the trispectrum and thus carry complementary information on the parity-odd signal. In particular, they are sensitive to different soft limits of the trispectrum. To empirically validate the effectiveness of these estimators, we tested them on a set of mock simulations with an injected parity-odd four-point correlation function. On these mocks, we compared the result of the estimator with semi-analytical theoretical calculations. The agreement between the data and our theoretical calculations demonstrates that our estimators robustly capture the injected parity-odd signal.

Beyond validation, we investigated the sensitivity of our new estimators by computing the correlation matrix on Gaussian realizations. We found that different bins of our estimator are predominantly uncorrelated. Furthermore, a $\chi^2$-squared test on simulated parity-violating mocks highlighted the substantial statistical significance of both estimators, with the scalar variant exhibiting enhanced sensitivity compared to its vector counterpart. This sensitivity is specific to the choice of the primordial parity-violating trispectrum template and filtering functions. In particular, we did not attempt to optimize the filtering functions by maximizing signal-to-noise. Such an optimization could dramatically increase the significance. We plan to explore this in future work.

Applying our estimator to realistic survey data requires several developments. Surveys observe galaxies within a specific survey mask and selection function. We must incorporate these into our estimator to remove the spurious clustering signals they would otherwise produce. Moreover, galaxies are discrete tracers, and this results in shot noise that should be characterized and subtracted from the POP spectra. We leave these developments for future work. After overcoming these challenges, we expect the POP estimators will emerge as an essential tool in the search for parity violation in current and future CMB and galaxy surveys.

\section*{Acknowledgements}
We thank Toshiki Kurita, Fabian Schmidt, Azadeh Moradinezhad Dizgah, Robert Cahn, Donghui Jeong, and Ue-Li Pen for helpful discussions. 
AC acknowledges funding support from the Initiative Physique des Infinis (IPI), a research training program of the Idex SUPER at Sorbonne Universit\'e.
This work was supported in part by the Excellence Cluster ORIGINS, which is funded by the Deutsche Forschungsgemeinschaft (DFG, German Research Foundation) under Germany’s Excellence Strategy: Grant No.~EXC-2094 - 390783311. 
JH acknowledges funding from the European Union's Horizon 2020 research and innovation programme under the Marie Skłodowska-Curie grant agreement no.\ 101007633. The Kavli IPMU is supported, in part, by World Premier International Research Center Initiative (WPI), MEXT, Japan. ZS thanks the Max Planck Institute for Astrophysics for hospitality during some of the period during which this work was performed.

\appendix

\section{The Binned Parity-Odd Trispectrum and the Binned POP Spectra}
\label{app:bin}

Measuring the parity-odd trispectrum from data requires choosing a binning scheme for the tetrahedral configurations of wave vectors. To form a bin of nearby tetrahedra, we specify the side lengths $k_1$, $k_2$, $k_3$ $k_4$, and the diagonal lengths $K$ and $\tilde{K}$. We then select the set of wave vectors satisfying
\begin{align}
    \vk_1 + \vk_2 + \vK =\ & 0 \, , \\
    \vk_3 + \vk_4 - \vK =\ & 0 \, , \\
    \vk_1 + \vk_4 + \tilde{\vec K} =\ & 0 \, .
\end{align}
We define the six-dimensional binning scheme through the integral
\begin{align}
    \int_{\mathbf{bin}} \equiv \int_{\vec{q}_1\in k_1} \int_{\vec{q}_2\in k_2} \int_{\vec{q}_3\in k_2} \int_{\vec{q}_4\in k_4} \Theta\Big(4\big(|\vq_1 + \vq_2| - K\big)^2 - \Delta k^2\Big) \,
    \Theta\Big(4\big(|\vq_1 + \vq_4| - \tilde{K}\big)^2 - \Delta k^2\Big) \, .
\end{align}
The integral notation is defined in Table~\ref{tab:intsum}. The Heaviside functions enforce $K-\Delta k / 2 \leq |\vq_1 + \vq_2| \leq K+\Delta k / 2$ and $\tilde{K}-\Delta k / 2 \leq |\vq_1 + \vq_4| \leq \tilde{K}+\Delta k / 2$, which defines the binning scheme for the diagonals.

We isolate the binned parity-odd trispectrum by integrating the product of the triple product and the four-point expectation value over the bin,
\begin{align}
    \bar{\tau}_-(\ka,\kb,\kc, \kd, K, \tilde{K}) =\ & \mathcal{N}_{\tau_-}^{-1} \displaystyle\int_{\mathbf{bin}} -i \vqa \cdot (\vqb \times \vqc) \, \Big\langle \Phi(\vqa) \Phi(\vqb) \Phi(\vqc) \Phi(\vqd) \Big\rangle \\     
    \label{eq:tbar1}
    =\ & \mathcal{N}_{\tau_-}^{-1} \displaystyle\int_{\mathbf{bin}}(2\pi)^3 \dd{\vqa+\vqb+\vqc+\vqd} \big[\vqa \cdot (\vqb \times \vqc)\big]^2 \tau_-\big(\qa, \qb, \qc, \qd, |\vqa+\vqb|, |\vqa+\vqd|\big) \, .
\end{align}
Here, $\mathcal{N}_{\tau_-}$ is a normalization factor, which we will define later. The minus sign in the first line is for convenience, and we could absorb it into the normalization factor. We split the Dirac delta function as follows:
\begin{align}
    \label{eq:ddd}
    (2\pi)^3\dd{\vqa+\vqb+\vqc+\vqd} = (2\pi)^9 \int\displaylimits_{\vQ,\tilde{\vQ}} \dd{\vqa + \vqb + \vQ} \dd{\vqc + \vqd - \vQ} \dd{\vqa + \vqd + \tilde{\vQ}} \, .
\end{align}
The right-hand side of this expression imposes the condition that the tetrahedron consists of two triangles, $\{\vqa, \vqb, -\vQ\}$ and  $\{\vqc, \vqd, \vQ\}$, which are joined along their common side of length $Q$. The vectors $\vqa$ and $\vqd$ are additionally constrained to form a triangle $\{\vqa, \vqd, -\tilde{\vQ}\}$. After substituting Eq.~\eqref{eq:ddd} into Eq.~\eqref{eq:tbar1}, the Heaviside functions limit the integrations over $Q$ and $\tilde{Q}$. Then, with the integral notation in Table~\ref{tab:intsum}, the binned parity-odd trispectrum is given by
\begin{align} \label{eq:sixbin}
    \bar{\tau}_-(k_1, k_2, k_3, k_4, K, \tilde{K}) = 
    \mathcal{N}_{\tau_-}^{-1} \int_{6D\mbox{-}\mathbf{bin}} \Big[ (2\pi)^9 \dd{\vq_1 + \vq_2 + \vec Q} \dd{\vq_3 + \vq_4 - \vec Q} \dd{\vq_1 + \vq_4 + \vec{\tilde{Q}}} \nonumber \\
    \times\, \big[\vq_1 \cdot (\vq_2 \times \vq_3)\big]^2 \,
    \tau_-(\qa, \qb, \qc, \qd, Q, \tilde{Q}) \Big] \, .
\end{align}

\begin{table}
    \centering
    \begin{tabular}{|c@{\hspace{2em}}|@{\hspace{2em}}l|}
        \hline
        \textbf{Notation} & \textbf{Definition} \\
        \hline
        $\displaystyle\int_{q \in k}$ & $\displaystyle\int_{k -\Delta k/2}^{k + \Delta k/2} \frac{k^2 \dee k}{2 \pi^2}$ \\
        \hline
        $\displaystyle\int_{\hat{q}}$ & $\displaystyle\int \frac{\dee \Omega_{\hat{q}}}{4\pi}$ \\
        \hline
        $\displaystyle\int_{\vq \in k}$ & $\displaystyle\int_{\hat{q}} \int_{q \in k}$ \\
        \hline
        $\displaystyle\int_{6D\mbox{-}\mathbf{bin}}$ & $\displaystyle\int_{\vec{q}_1\in k_1} \int_{\vec{q}_2\in k_2} \int_{\vec{q}_3\in k_2} \int_{\vec{q}_4\in k_4} \int_{\vec{Q} \in K} \int_{\vec{\tilde{Q}} \in \tilde{K}}$ \\
        \hline
        $\displaystyle\int_{5D\mbox{-}\mathbf{bin}}$ & $\displaystyle\int_{\vq_1\in k_1} \int_{\vq_2 \in k_2} \int_{\vq_3\in k_3} \int_{\vq_4\in k_4} \int_{\vec{Q} \in K}$ \\
        \hline
        $\displaystyle\int_{5D\mbox{-}{\rm bin}}$ & $\displaystyle\int_{q_1\in k_1} \int_{q_2\in k_2} \int_{q_3\in k_3} \int_{q_4 \in k_4} \int_{Q \in K}$ \\
        \hline
        $\displaystyle\sum_{\Li\Mi}$ & $\displaystyle
        \sum_{L_1=0}^{\infty} \sum_{M_1=-L_1}^{L_1}
        \sum_{L_2=0}^{\infty} \sum_{M_2=-L_2}^{L_2} 
        \sum_{L_3=0}^{\infty} \sum_{M_3=-L_3}^{L_3}$ \\
        \hline
    \end{tabular}
    \caption{Fourier-space integral and harmonic-space summation notation. The notation $q\in k$ indicates a wave vector magnitude belonging to a bin centered on $k$ with width $\Delta k$. The integral over $\dee \Omega_{\hat{q}}$ is the solid angle integral over the azimuthal and polar angles associated with the wave vector $\vq$ in polar coordinates.}
    \label{tab:intsum}
\end{table}

Unfortunately, the full six-dimensional trispectrum is difficult to analyze directly. The first two Dirac delta functions factor into integrals over two triangles, but the third couples these together, making the whole integral not factorizable. For this reason, we extend the $\tilde{Q}$ bin to be unrestricted, $\tilde{Q}\in[0,\infty)$, which fully averages over the second diagonal. An alternative approach is to implement isotropic basis functions, which discretize the angular dependence through spherical harmonic expansions \citep{Cahn:2020axu}, but we will not pursue that here.

The five-dimensional binned trispectrum, with notation from Table~\ref{tab:intsum}, is
\begin{align}
    \label{eq:tbar}
    \bar{\tau}_-(k_1, k_2, k_3, k_4, K) =\ &
    \mathcal{N}_{\tau_-}^{-1} \int_{5D\mbox{-}\mathbf{bin}}(2\pi)^6 \dd{\vq_1 + \vq_2 + \vec Q} \dd{\vq_3 + \vq_4 - \vec Q} \,
    \big[\vq_1 \cdot (\vq_2 \times \vq_3)\big]^2 \, \tau_-\big(q_1, q_2, q_3, q_4, Q, |\vq_1 + \vq_4|\big) \, ,
\end{align}
The third Dirac delta function in Eq.~\eqref{eq:sixbin} enforced $\tilde{Q}=|\vq_1 + \vq_4|$. We choose the somewhat arbitrary normalization factor $\mathcal{N}_{\tau_-}$ to be
\begin{align}
    \mathcal{N}_{\tau_-} = \int_{5D\mbox{-}\mathbf{bin}} (2\pi)^9 \dd{\vq_1 + \vq_2 + \vec Q} \dd{\vq_3 + \vq_4 - \vec Q} \,
    \big[\vq_1 \cdot (\vq_2 \times \vq_3)\big]^2 \, ,
\end{align}
so Eq.~\eqref{eq:tbar} is the weighted average of $\tau_-(q_1, q_2, q_3, q_4, Q, |\vqa+\vqd|)$ within the bin.

Expressing both Dirac delta functions as the Fourier transform of plane waves enables us to compute most angular integrals analytically. The calculation simplifies greatly by noticing that we can rewrite the triple product as $\vq_1 \cdot (\vq_2 \times \vq_3) = -\vec Q \cdot (\vq_1 \times \vq_3)$, which is more symmetric between the two triangles forming the tetrahedron. However, if the trispectrum depends on the diagonals, the angular dependence on $\vq_1 \cdot \vq_4$ may prevent us from analytically computing all angular integrals. From here on, we will restrict to the case where the trispectrum has no explicit dependence on the diagonals.
		
The spherical harmonic expansion of the squared triple product is
\begin{align}
    \left[ \vQ\cdot(\vqa\times\vqc)\right]^2 = -6(4\pi)^3 (Q\qa\qc)^2 \sum_{L_i,M_i} \DLM \Ya{\hqa}  \Yb{\hqc}  \Yc{\hQ} \, ,
\end{align}
where the coefficients are
\begin{align}
    \DLM = \tjabc \njabc \prod_{i=1}^{3}\sqrt{\frac{2\Li+1}{4\pi}} \tji \, .
\end{align}
Here, the 2 by 3 matrices are Wigner 3-$j$ symbols, and the 3 by 3 matrix is a Wigner 9-$j$ symbol. The form of the coefficients in the product restricts $\Li\in\{0,2\}$.

The trispectrum estimator is an integral over tetrahedra formed by two triangles: $\{\vqa, \vqb, -\vQ\}$ and $\{\vqc, \vqd, \vQ\}$. By taking one vector from each triangle and the common vector shared by both triangles in the triple product, we preserve the symmetry between the two pairs ($\vqa$,$\vqb$) and ($\vqc$,$\vqd$), greatly simplifying the calculation. Each triangle contributes a Dirac delta function and a $Y_{LM}$ from a vector that is not $\pm\vQ$. Expressing the Dirac delta functions as Fourier transforms of plane waves allows us to compute all angular integrals,  
\begin{align}
    \int\displaylimits_{\vqa\in\ka,\vqb\in\kb} (2\pi)^3 \dd{\vqa+\vqb-\vQ} \Ya{\hqa} =\ &
    \Ya{\hQ} \int\displaylimits_{\qa\in\ka,\qb\in\kb}
    4\pi \int_{0}^{\infty} \!\! \dee x ~ x^2 \, \ja{\qa x} \jo{\qb x}  \ja{Q x} \\
    =\ & \Ya{\hQ} \int\displaylimits_{\qa\in\ka,\qb\in\kb}
     I_{\La 0 \La}(\qa, \qb, Q) \, ,
\end{align}
and similarly,
\begin{align}
    \int\displaylimits_{\vqc\in\kc,\vqd\in\kd} (2\pi)^3 \dd{\vqc+\vqd+\vQ} \Yb{\hqc} =\ &
    (-1)^{\Lb}\Yb{\hQ} \hspace{-16pt} \int\displaylimits_{\qc\in\kc,\qd\in\kd} 
    4\pi \int_0^{\infty} \!\!\dee x ~ x^2 \, \jb{\qc x} \jo{\qd x}  \jb{Q x} \\
    =\ & (-1)^{\Lb} \Yb{\hQ} \int\displaylimits_{\qc\in\kc,\qd\in\kd} I_{\Lb 0 \Lb}(\qc, \qd, Q) \, .
\end{align}
Since $\Lb\in\{0,2\}$, we have $(-1)^{\Lb}=1$, so we can drop this factor. Both triangles contribute factors to the integrand that have the same form, illustrating the symmetric roles of the two triangles that form the tetrahedron. The triple-spherical Bessel integrals are given by
\begin{align}
    I_{\La 0 \La}(\qa, \qb, Q) \equiv \ & 4\pi \int_0^{\infty} \!\! \dee x ~ x^2 \, \ja{\qa x} \jo{\qb x}  \ja{Q x} \\
    = \ & \frac{\pi^2}{\qa\qb Q} \Theta\big(\sin^2\!\theta_{12}\big) P_{\La}\big(\cos \theta_{12}\big) \, .
\end{align}
Here, $\theta_{12}$ is the angle between $\vqa$ and $\vQ$. As discussed in the main text, the Heaviside function enforces the triangle inequalities that ensure $\vqa$, $\vqb$, and $-\vQ$ can form a closed triangle. $P_{\La}$ is the $\La^{\mathrm{th}}$ Legendre polynomial. The only relevant ones will be
\begin{align}
    P_0(x) = 1 \, ,\quad
    P_2(x) = \frac{1}{2}\lb 3 x^2 - 1 \rb \, .
\end{align}
At this point, three $Y_{\Li\Mi}$ products remain, all with the argument $\hQ$. Integrating over $\hQ$ gives Gaunt's integral (divided by $4\pi$),\footnote{\url{https://dlmf.nist.gov/34.3}, Eq. 34.3.22}
\begin{align}
    \GLM & = \int_{\hQ} \Ya{\hQ}  \Yb{\hQ} \Yc{\hQ} \\
            & = \tjabc \tjabco \prod_{i=1}^{3} \sqrt{\frac{2\Li+1}{4\pi}} \, .
\end{align}

The form of the trispectrum after computing all angular integrals is
\begin{align}
    \bar{\tau}_{-}(\ka,\kb,\kc,\kd,K) = -6 \, \mathcal{N}_{\tau_-}^{-1} \int_{5D\mbox{-}{\rm bin}} (Q \qa \qc)^2 \sum_{\Li\Mi} \DLM \GLM   I_{\La 0 \La}(\qa, \qb, Q) I_{\Lb 0 \Lb}(\qc, \qd, Q) \, \tau_{-}(\qa,\qb,\qc,\qd) \, ,
\end{align}
where the summation notation is defined in Table~\ref{tab:intsum}. Using the orthogonality relation for the 3-$j$ symbol\footnote{\url{https://dlmf.nist.gov/34.3} Eq. 34.3.18}
\begin{align}
\sum_{M_1=-L_1}^{L_1} \sum_{M_2=-L_2}^{L_2} \sum_{M_3=-L_3}^{L_3} \tjabc^2 = 1 \, ,
\end{align}
we can perform the sums over all $\Mi$. Since neither triple-spherical Bessel integral involves $\Lc$, we can also sum over it, which gives
\begin{align}
        \bar{\tau}_{-}(\ka,\kb,\kc,\kd,K) = 6 \,
        \mathcal{N}_{\tau_-}^{-1} \int_{5D\mbox{-}{\rm bin}} (Q \qa \qc)^2 \sum_{\La,\Lb} \HL I_{\La 0 \La}(\qa, \qb, Q) I_{\Lb 0 \Lb}(\qc, \qd, Q) \tau_{-}(\qa,\qb,\qc,\qd) \, ,
\end{align}
where the coefficients now simplify considerably:
\begin{align}
    \HL & = -\sum_{\Lc} \tjabco \njabc \prod_{i=1}^{3} (2\Li+1) \tji \\
           & = \frac{i^{\La+\Lb}}{27} \, .
\end{align}	
The two remaining sums factorize,
\begin{align}
    \bar{\tau}_{-}(\ka,\kb,\kc,\kd,K) = \frac{2}{9} \mathcal{N}_{\tau_-}^{-1} \int_{5D\mbox{-}{\rm bin}} (Q \qa \qc)^2 \Big[ \sum_{\La} i^{\La} I_{\La 0 \La}(\qa, \qb, Q) \Big] \Big[ \sum_{\Lb} i^{\Lb} I_{\Lb 0 \Lb}(\qc, \qd, Q) \Big]  \tau_{-}(\qa,\qb,\qc,\qd) \, .
\end{align}
Each sum only has two terms, so we can sum them explicitly and expand the Legendre polynomials,
\begin{align}
    \bar{\tau}_-(\ka,\kb,\kc,\kd,K) = \frac{\pi^4}{2} \mathcal{N}_{\tau_-}^{-1} \int_{5D\mbox{-}{\rm bin}}
    \frac{\qa\qc}{\qb\qd} \, \Theta\big(\sin^2\!\theta_{12}\big)
    \sin^2\!\theta_{12} \, \Theta\big(\sin^2\!\theta_{34}\big) 
    \sin^2\!\theta_{34} \, \tau_{-}(\qa,\qb,\qc,\qd) \, ,
\end{align}
where we set $\tau_-=1$ to obtain the normalization factor,
\begin{align}
    \mathcal{N}_{\tau_-}  = \frac{\pi^4}{2} \int_{5D\mbox{-}{\rm bin}}
    \frac{\qa\qc}{\qb\qd} \Theta\big(\sin^2\!\theta_{12}\big)
    \sin^2\!\theta_{12} \Theta\big(\sin^2\!\theta_{34}\big) 
    \sin^2\!\theta_{34} \, .
\end{align}

Instead of restricting the integration bounds, we could define filter functions with Heaviside functions that impose the binning. Then, the binned parity-odd trispectrum takes the form,
\begin{align}
    \label{eq:tbin}
    \bar{\tau}_-(\ka,\kb,\kc,\kd,K) = \frac{\pi^4}{2} \mathcal{N}_{\tau_-}^{-1} \int\displaylimits_{\qa,\qb,\qc,\qd,Q} \Big [ &
    \frac{\qa\qc}{\qb\qd} \, \Theta\big(\sin^2\!\theta_{12}\big)
    \sin^2\!\theta_{12} \, \Theta\big(\sin^2\!\theta_{34}\big) 
    \sin^2\!\theta_{34} \nonumber \\
    & \times \, f_a(\qa) f_b(\qb) f_c(\qc) f_d(\qd) f_e(Q) \, \tau_{-}(\qa,\qb,\qc,\qd) \Big] \, .
\end{align}
By taking all but $f_e$ to be generic filter functions, rather than just Heaviside functions selecting spherical shells, and taking the thin-bin limit of $f_e$, this has the same form as the vector POP spectrum in Eq.~\eqref{eq:vpvth}. Similarly, by taking all but $f_a$ to be generic filter functions and taking the thin-bin limit of $f_a$, this has the same form as the scalar POP spectrum in Eq.~\eqref{eq:spsth}. Thus, computing a POP spectrum is equivalent to computing a one-dimensional subset of bins for the full five-dimensional trispectrum.

\bibliographystyle{mnras}
\bibliography{refs.bib}

\label{lastpage}

\end{document}